\documentclass{pasj00}
\draft
\usepackage{lscape}

\begin{document}
\SetRunningHead{Tatematsu et al.}{Chemical Variation in Orion. II.}
\Received{2013/06/17}
\Accepted{2013/09/19}

\title{Chemical Variation in Molecular Cloud Cores in the Orion A Cloud. II.}

\author{Ken'ichi \textsc{Tatematsu},\altaffilmark{1,2}
Satoshi \textsc{Ohashi},\altaffilmark{3}
Tomofumi \textsc{Umemoto},\altaffilmark{1,2}
Jeong-Eun \textsc{Lee},\altaffilmark{4}
Tomoya \textsc{Hirota},\altaffilmark{1,2}
Satoshi \textsc{Yamamoto},\altaffilmark{5}
Minho \textsc{Choi},\altaffilmark{6}
Ryo \textsc{Kandori},\altaffilmark{1}
and
Norikazu \textsc{Mizuno}\altaffilmark{1,3}
}
\altaffiltext{1}{National Astronomical Observatory of Japan, 
2-21-1 Osawa, Mitaka, Tokyo 181-8588}
\altaffiltext{2}{Department of Astronomical Science, 
The Graduate University for Advanced Studies (Sokendai), 
2-21-1 Osawa, Mitaka, Tokyo 181-8588}
\altaffiltext{3}{Department of Astronomy, The University of Tokyo, Bunkyo-ku, Tokyo 113-0033}
\altaffiltext{4}{School of Space Research, Kyung Hee University, 
Seocheon-Dong, Giheung-Gu, Yongin-Si, Gyeonggi-Do, 446-701, South Korea}
\altaffiltext{5}{Department of Physics, The University of Tokyo, Bunkyo-ku, Tokyo 113-0033}
\altaffiltext{6}{Korea Astronomy and Space Science Institute, 
Daedeokdaero 776, Yuseong, Daejeon 305-348, South
Korea}

\email{k.tatematsu@nao.ac.jp, satoshi.ohashi@nao.ac.jp, umemoto.tomofumi@nao.ac.jp, 
jeongeun.lee@khu.ac.kr,
tomoya.hirota@nao.ac.jp,
yamamoto@taurus.phys.s.u-tokyo.ac.jp,
minho@kasi.re.kr,
r.kandori@nao.ac.jp,
norikazu.mizuno@nao.ac.jp
}


%

\KeyWords{ISM: clouds
---ISM: individual (Orion Nebula, Orion Molecular Cloud)
---ISM: molecules
---ISM: structure---stars: formation} 

\maketitle

\begin{abstract}
We have mapped six molecular cloud cores 
in the Orion A giant molecular cloud (GMC),
whose kinetic temperatures range from 10 to 30 K,
in CCS and N$_2$H$^+$ with Nobeyama 45 m radio telescope to study
their chemical characteristics.
We identified 31 intensity peaks in the CCS and N$_2$H$^+$ emission
in these molecular cloud cores.
It is found for cores with temperatures lower than $\sim$ 25 K
that 
the column density ratio of $N$(N$_2$H$^+$)/$N$(CCS) is
low toward starless core regions while it is high
toward star-forming core regions, in case that we detected both of 
the CCS and N$_2$H$^+$ emission.
This is very similar to the tendency found in dark clouds 
(kinetic temperature $\sim$ 10 K).
The criterion found in the Orion A GMC is $N$(N$_2$H$^+$)/$N$(CCS) $\sim 2-3$.
In some cases, the CCS emission is detected toward protostars as well as 
the N$_2$H$^+$ emission.
Secondary late-stage CCS peak in the chemical evolution caused by CO 
depletion may be 
a possible explanation for this.
We found that the chemical variation of CCS and N$_2$H$^+$ can also 
be used as a tracer of evolution in warm (10$-$25 K) GMC cores.
On the other hand, some protostars do not accompany N$_2$H$^+$ intensity peaks
but are associated with dust continuum emitting regions, 
suggesting that the N$_2$H$^+$ abundance might be decreased 
due to CO evaporation in warmer star-forming sites.
\end{abstract}

\section{Introduction}

In nearby dark clouds, molecules such as CCS, HC$_3$N, NH$_3$, and N$_2$H$^+$,
and the neutral carbon atom C$^0$
are known to be good tracers of the chemical evolution
(e.g., \cite{hir92,suz92,ben98,mae99,lai00,hir02,hir09}).
The carbon-chain molecules, CCS and HC$_3$N trace 
the early chemical 
evolutionary stage, whereas N-bearing molecules, 
NH$_3$ and N$_2$H$^+$ trace
the late stage.
On the other hand, the chemical evolution of molecular cloud cores 
in giant molecular clouds (GMCs) is less understood,
compared with that of nearby dark cloud cores.
Most of stars in the Galaxy form in GMCs, 
and then the chemical evolution of molecular cloud cores 
in them
is of our great interest to study the star forming process in the Galaxy.
We wonder how different the chemical properties of molecular cloud cores
in GMCs are, compared with those in nearby dark clouds.
Dark clouds and GMCs are different in their physical properties
and associated star formation
(e.g., \cite{tur88}).
GMCs are larger, more massive and warmer than dark clouds.
Dark clouds show only isolated low-mass star formation, but GMCs show
massive star formation and formation of star clusters.

The Orion A cloud is an archetypal GMC.
\citet{ung97} studied the region near Orion KL in many molecular lines, 
and have shown difference in the molecular abundances among 
the Orion KL region, 
the Orion Bar, and the molecular ridge.
\citet{tat93a} mapped the Orion A giant molecular cloud
in CS $J$ = 1$-$0
and have shown the chemical variation between
Orion KL and OMC-2 
by comparing the distribution of CS with that of NH$_3$.
\citet{tat08} showed that
the N$_2$H$^+$ emission is widely distributed over the $\int$-shaped filament of
the Orion A GMC, and
pointed out that there is 2 arcmin (0.3 pc)-scale displacement between
the HC$_3$N and N$_2$H$^+$ distribution near X-ray emitting protostars in
the OMC-3 region at the northern end of the Orion A GMC.
\citet{tat10} made single-point observations toward the cloud core centers
in the Orion A cloud
cataloged by \citet{tat93a}, and detected the CCS emission in $\sim$ 30\% of them.  
They discussed the chemical variation in this cloud on the basis of the column density ratio
of NH$_3$ to CCS 
and that of DNC to HN$^{13}$C.  They found that the NH$_3$/CCS and DNC/HN$^{13}$C
column density ratios are lower toward regions with warmer temperatures and that
the NH$_3$/CCS column density ratio shows a global variation along the Orion A GMC filament.
\citet{joh03} compared submillimeter dust continuum sources
in the Orion A GMC with the line emission from other molecules such
as CO, H$_2$CO, and CH$_3$OH.

In this study, we investigate the chemical characteristics of
molecular cloud cores in the Orion A GMC
through new mapping observations of CCS and N$_2$H$^+$.
The purpose of this study is to investigate whether 
our knowledge of the chemical evolution found 
in cold dark clouds 
(with kinetic temperatures $T_k \sim$10 K; see, e.g., \cite{ben98}, \cite{hir09} and references therein)
is valid also for cores in Orion A GMC ($T_k \sim$ 10$-$30 K)
or not.

The distance to the Orion A cloud is estimated to be 418 pc 
\citep{kim08}. At this distance,
1 arcmin corresponds to 0.12 pc.

\newpage

\section{Observations}

Observations were carried out by using the 45 m radio telescope
of Nobeyama Radio Observatory\footnote{Nobeyama Radio Observatory 
is a branch of the National Astronomical Observatory of Japan, 
National Institutes of Natural Sciences.} from 2013 January 13 to 24.  
The employed receiver front end was
the single-beam two-polarization two-sideband-separation (2SB) SIS receiver ``TZ1''.
We observed 
CCS $J_N$ = 7$_6-6_5$ at 81.505208 GHz \citep{hir06}
in both of the linear polarizations
and
N$_2$H$^+$ $J$ = 1$-$0 at 93.1737767GHz \citep{cas95}
in the V (vertical) polarization,
simultaneously.
The upper energy level $E_u$ for these transitions
are 15.3 and 4.3 K, respectively.
The FWHM beam size and main-beam efficiency $\eta_{MB}$ of the telescope
were 19.1$\pm$0.3 arcsec and 41$\pm$3\% at 86 GHz, respectively.
The receiver back end was the digital spectrometer ``SAM45''.
The spectral resolution 
was 30.52 kHz  (corresponding to $\sim$ 0.1 km s$^{-1}$).
Spectra were obtained at spacings of 20 arcsec and in the position-switching mode.
The employed off position was ($\Delta$ R. A., $\Delta$ Dec.)
= ($-$30$\arcmin$, 0$\arcmin$)
with respect to the map center taken from the core catalog of \citet{tat93a}.
The observed intensity is reported in terms of the corrected
antenna temperature $T_A^*$.
To derive the physical parameters, we use the main-beam 
radiation temperature $T_R$ = $T_A^*$/$\eta_{MB}$.
The telescope pointing was established by observing Orion KL
in the 43-GHz SiO maser line every 60$-$80 min.

We selected 14 cores toward which 
CCS $J_N$ = 4$_3-3_2$ at 45.379033 GHz
was detected in the single pointing observation of \citet{tat10}. 
The upper energy level $E_u$ for CCS $J_N$ = 4$_3-3_2$
is 5.4 K.
The selected cores are TUKH003, 021, 040, 049, 056, 059, 069, 083, 088, 097, 104,
105, 117, and 122.
Table 1 lists the map centers and employed off positions.
First, we have made observations toward 3$\times$3 positions
at spacings of 20 arcsec 
centered at each of the 14 cores.
Out of the 14 observed cores, we have detected the 82 GHz CCS emission in 7 cores
(TUKH003, 021, 088, 097, 105, 117, and 122).
Out of these 7 cores, we have mapped larger area for 6 cores
also at spacings of 20 arcsec.
A larger map was not carried out for the remaining core TUKH105 due to limited observing time.
The average rms noise level in the 82 GHz CCS observations toward 14 cores
are 0.057$\pm$0.012 K at 30.52 kHz resolution (the rms noise level for maps of
each core will be shown later).
That in the N$_2$H$^+$ observations is 
about $\sqrt 2$ higher because the observations were made in a single polarization. 
Figure 1 shows the location of the six mapped cores illustrated on the CS
$J$ = 1$-$0 map (\cite{tat93a} and additional data, http://alma.mtk.nao.ac.jp/$\sim$kt/fits.html).

\begin{figure}
  \begin{center}

    \FigureFile(150mm,150mm){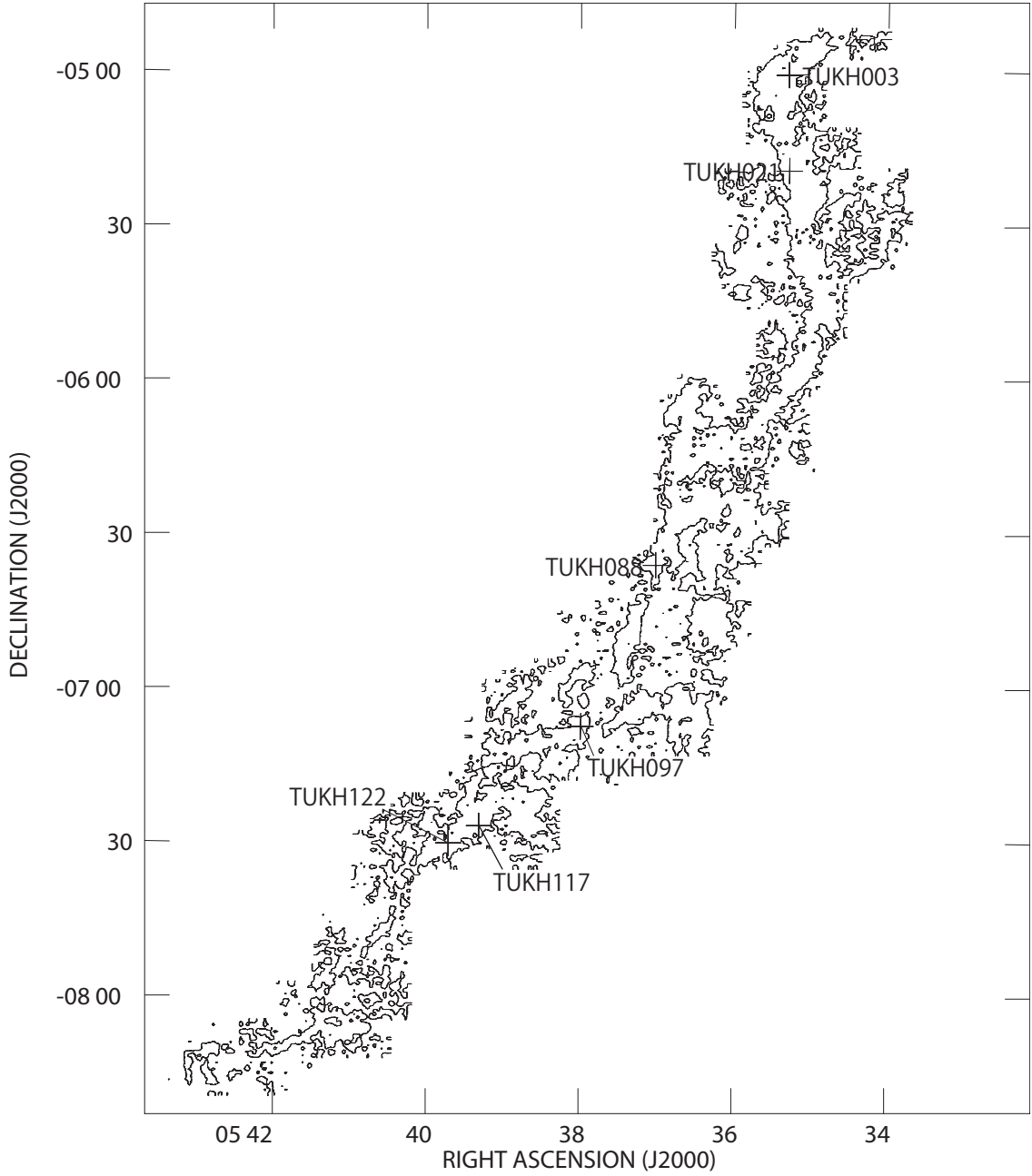}

  \end{center}
  \caption{
The locations of the six cores mapped in the present study.
The contour is 1.02 K (3$\sigma$) of the CS $J$ = 1$-$0 peak intensity map
(\cite{tat93a} and additional data, http://alma.mtk.nao.ac.jp/$\sim$kt/fits.html).
}\label{fig:figure1}
\end{figure}

The observed data were reduced by using the software package ``NewStar''
of Nobeyama Radio Observatory.

\section{Results and Discussion}

\subsection{Line Intensities and Their Variations}

Figures 2 to 7 show
the CCS $J_N$ = 7$_6-6_5$ contour maps superimposed
on the N$_2$H$^+$ $J$ = 1$-$0 gray-scale maps.
We plot protostars and their candidates taken from 
the Spitzer YSO (young stellar object) catalog 
of the Orion A and B GMCs based on
IRAC and MIPS observations \citep{meg12}. 
These authors carried out the protostar identification based on the 
colors constructed from IRAC and MIPS data and classified them into 
protostars (P), faint candidate protostars (FP), and red candidate protostars (RP,
objects that show MIPS detections at 24 $\mu$m but no detections at the 4.5, 5.8 or 8 $\mu$m IRAC bands). 
These sources are thought to be
Class 0, I, or flat spectrum
(with the spectral index determined between the 4.5, 5.8, and 8 $\mu$m IRAC bands and the 24 $\mu$m MIPS band; 
\cite{meg12} and references therein).

In the core regions shown in Figures 2$-$7,
there is no red candidate protostar and one faint candidate protostar, and the others are categorized as protostars.
\citet{meg12} assigned the MIPS magnitudes to IRAC point sources if they were
separated by $\leq$ 2.5 arcsec. 
Then, we assume that the typical accuracy in the coordinates of the protostars
is within 2.5 arcsec.
Table 2 lists the properties of protostars and one faint candidate protostar in regions
of Figures 2$-$7.
Other names are taken from \citet{eva86}, \citet{ang92}, \citet{chi97}, \citet{lis98},
\citet{tsu01}, \citet{tak09}, and \citet{tak13}.

\begin{figure}
  \begin{center}

    \FigureFile(150mm,150mm){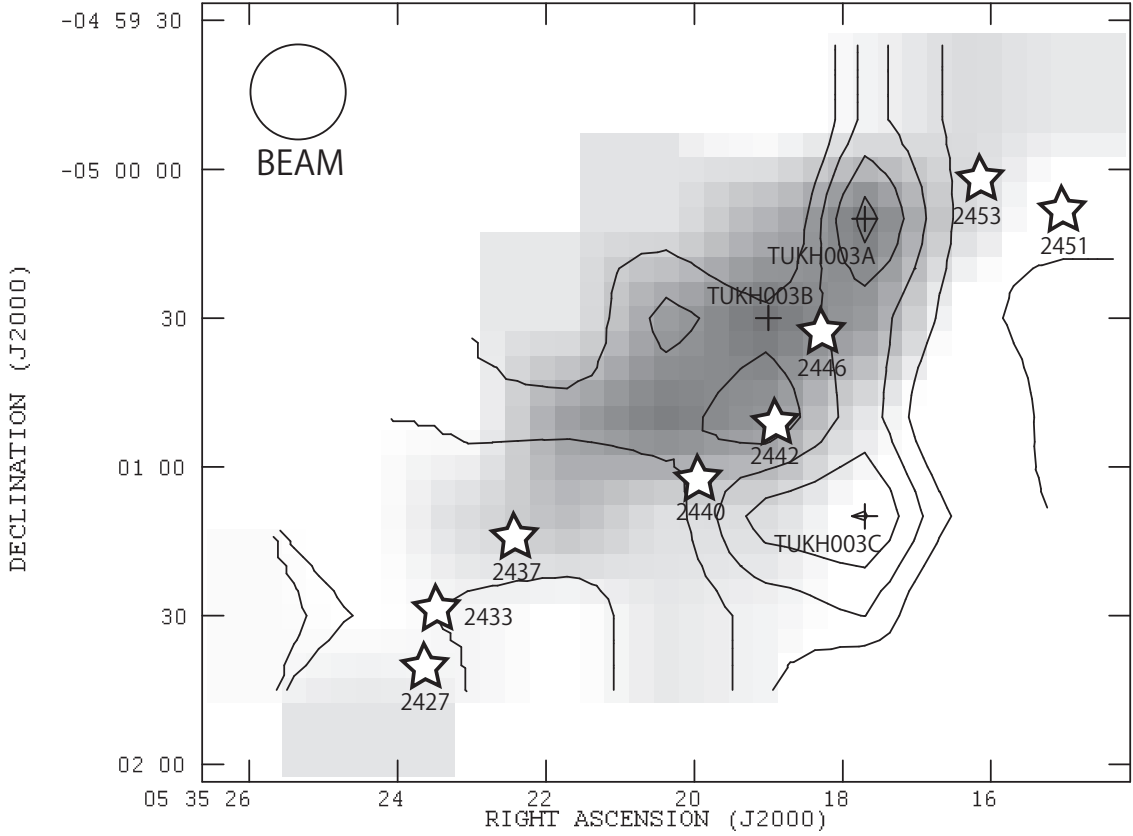}

  \end{center}
  \caption{
The CCS $J_N$ = 7$_6-6_5$ velocity-integrated intensity map
is superimposed
on the gray-scale map of the integrated intensity of the main
hyperfine component group
N$_2$H$^+$ $J$ = 1$-$0 $F_1$ = 2$-$1 for
TUKH003.
The velocity range for integration is 9.0 to 13.7 km s$^{-1}$.
The lowest contour is 3$\sigma$,
and the contour step is 1$\sigma$.
The 1$\sigma$ for the contour is 
2.7$\times$10$^{-2}$ K km s$^{-1}$.
The maximum in the gray scale corresponds to 
5.9 K km s$^{-1}$.
The open star signs represent the location of the protostar
in \citet{meg12}, and the associated numbers are their identification number.
The plus signs represent the location of our cataloged core intensity peak.
}\label{fig:figure2}
\end{figure}

\begin{figure}
  \begin{center}
    \FigureFile(150mm,150mm){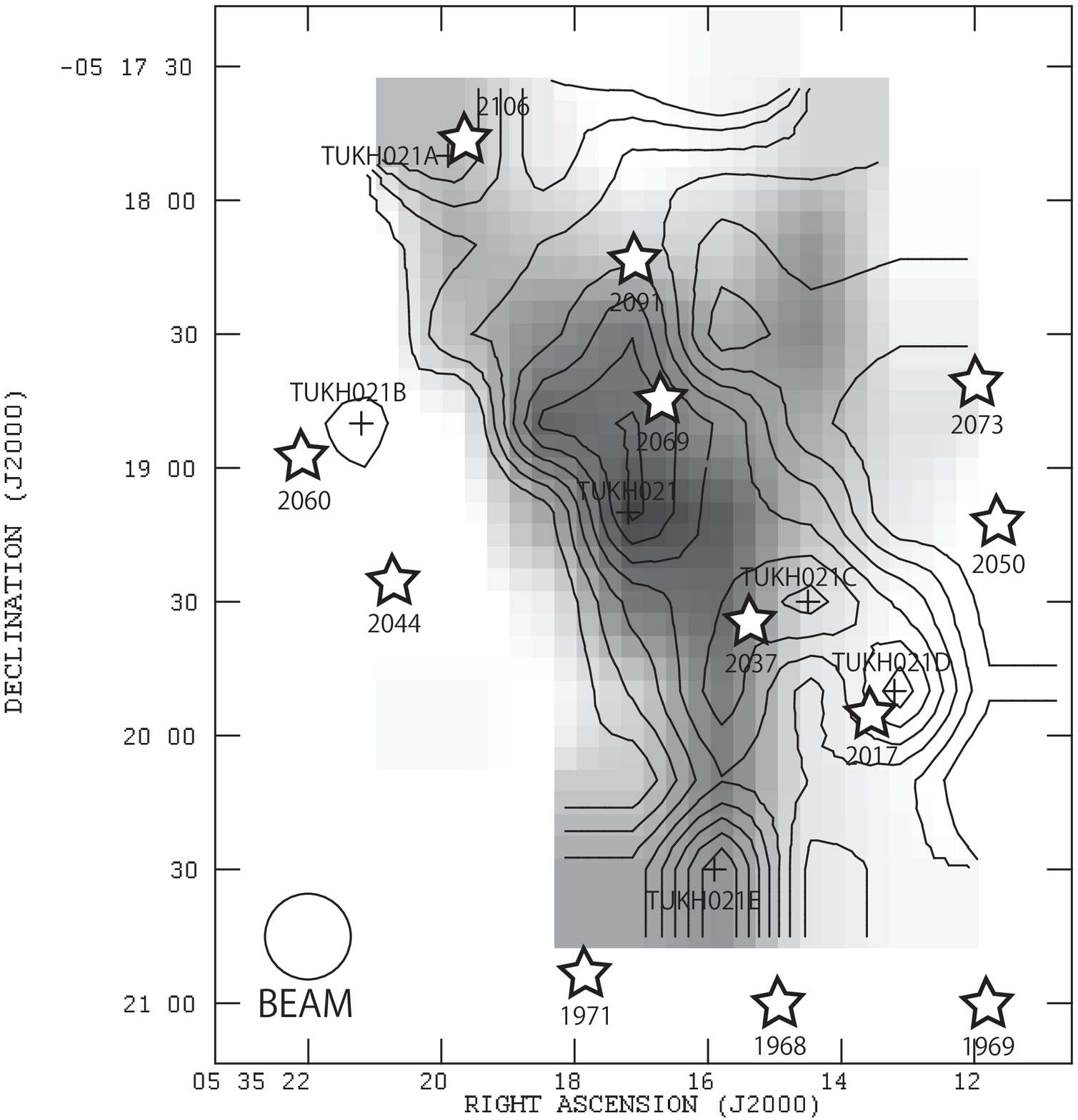}
  \end{center}
  \caption{The same as Figure 2 but for TUKH021.
The velocity range for integration is 7.0 to 12.3 km s$^{-1}$.
The 1$\sigma$ for CCS is 
4.6$\times$10$^{-2}$ K km s$^{-1}$,
and the maximum in the gray scale corresponds to 
12.5 K km s$^{-1}$.
}\label{fig:figure3}
\end{figure}

\begin{figure}
  \begin{center}
    \FigureFile(100mm,100mm){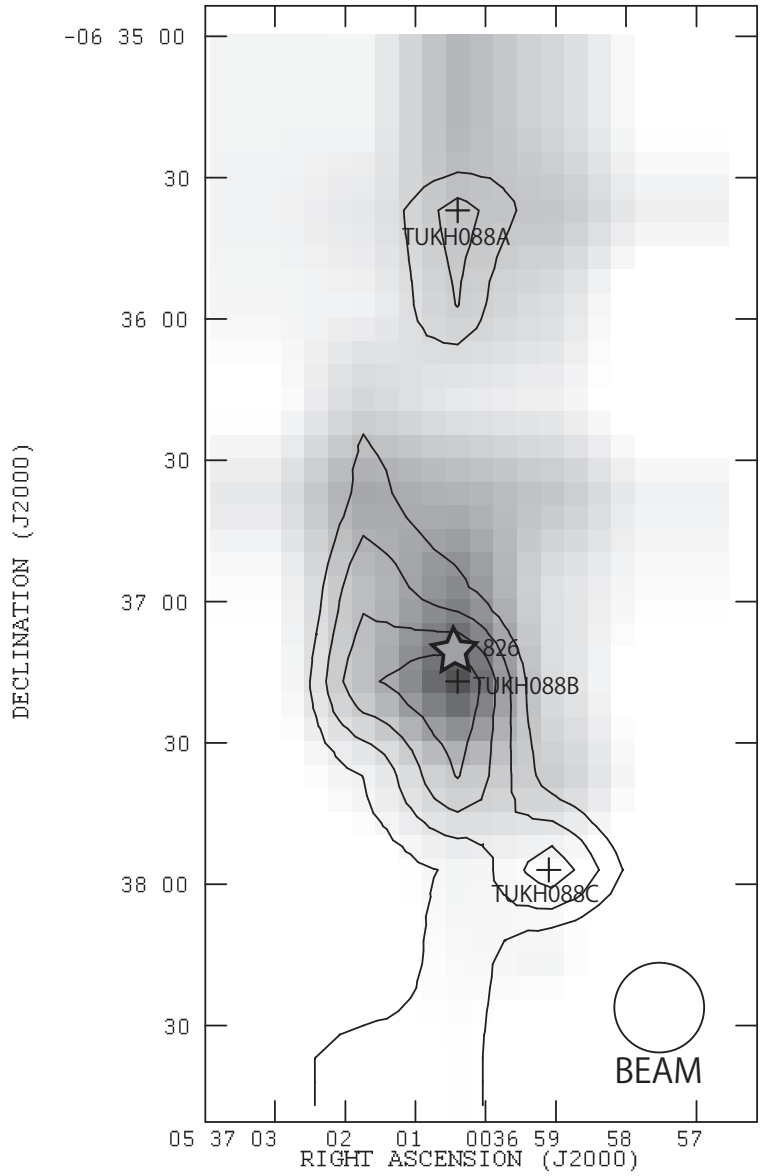}
  \end{center}
  \caption{The same as Figure 2 but for TUKH088.
The velocity range for integration is 5.0 to 8.2 km s$^{-1}$.
The 1$\sigma$ for CCS is 
3.4$\times$10$^{-2}$ K km s$^{-1}$,
and the maximum in the gray scale corresponds to 
1.33 K km s$^{-1}$.
The gray star sign represents the location of the faint candidate protostar
in \citet{meg12}.
}\label{fig:figure4}
\end{figure}

\begin{figure}
  \begin{center}

    \FigureFile(150mm,150mm){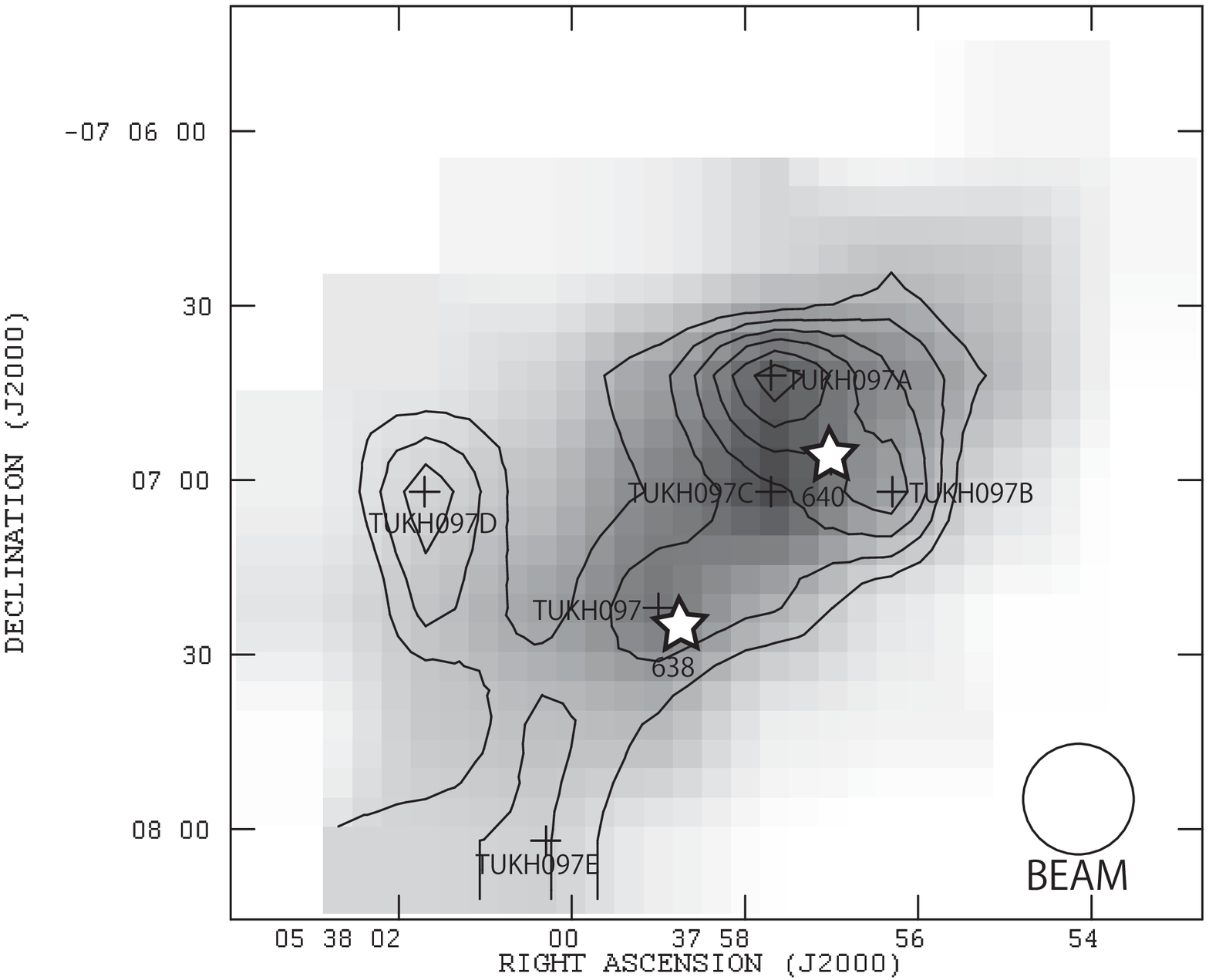}

  \end{center}
  \caption{The same as Figure 2 but for TUKH097.
The velocity range for integration is 4.3 to 8.0 km s$^{-1}$.
The 1$\sigma$ for CCS is 
3.1$\times$10$^{-2}$ K km s$^{-1}$,
and the maximum in the gray scale corresponds to 
2.75 K km s$^{-1}$.
}\label{fig:figure5}
\end{figure}

\begin{figure}
  \begin{center}
    \FigureFile(150mm,150mm){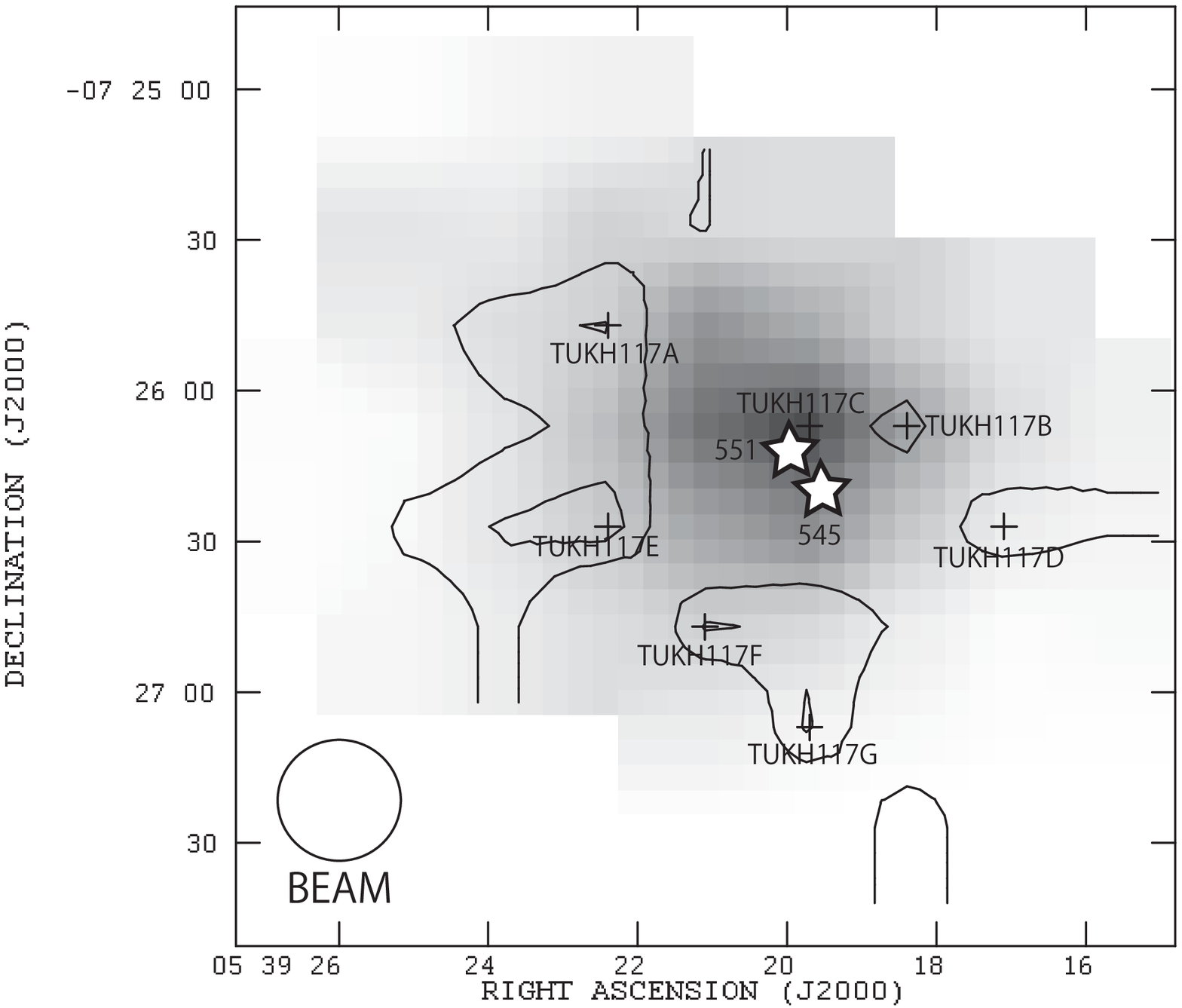}
  \end{center}
  \caption{The same as Figure 2 but for TUKH117.
The velocity range for integration is 3.0 to 7.5 km s$^{-1}$.
The 1$\sigma$ for CCS is 
3.7$\times$10$^{-2}$ K km s$^{-1}$,
and the maximum in the gray scale corresponds to 
5.06 K km s$^{-1}$.
}\label{fig:figure6}
\end{figure}

\begin{figure}
  \begin{center}
    \FigureFile(150mm,150mm){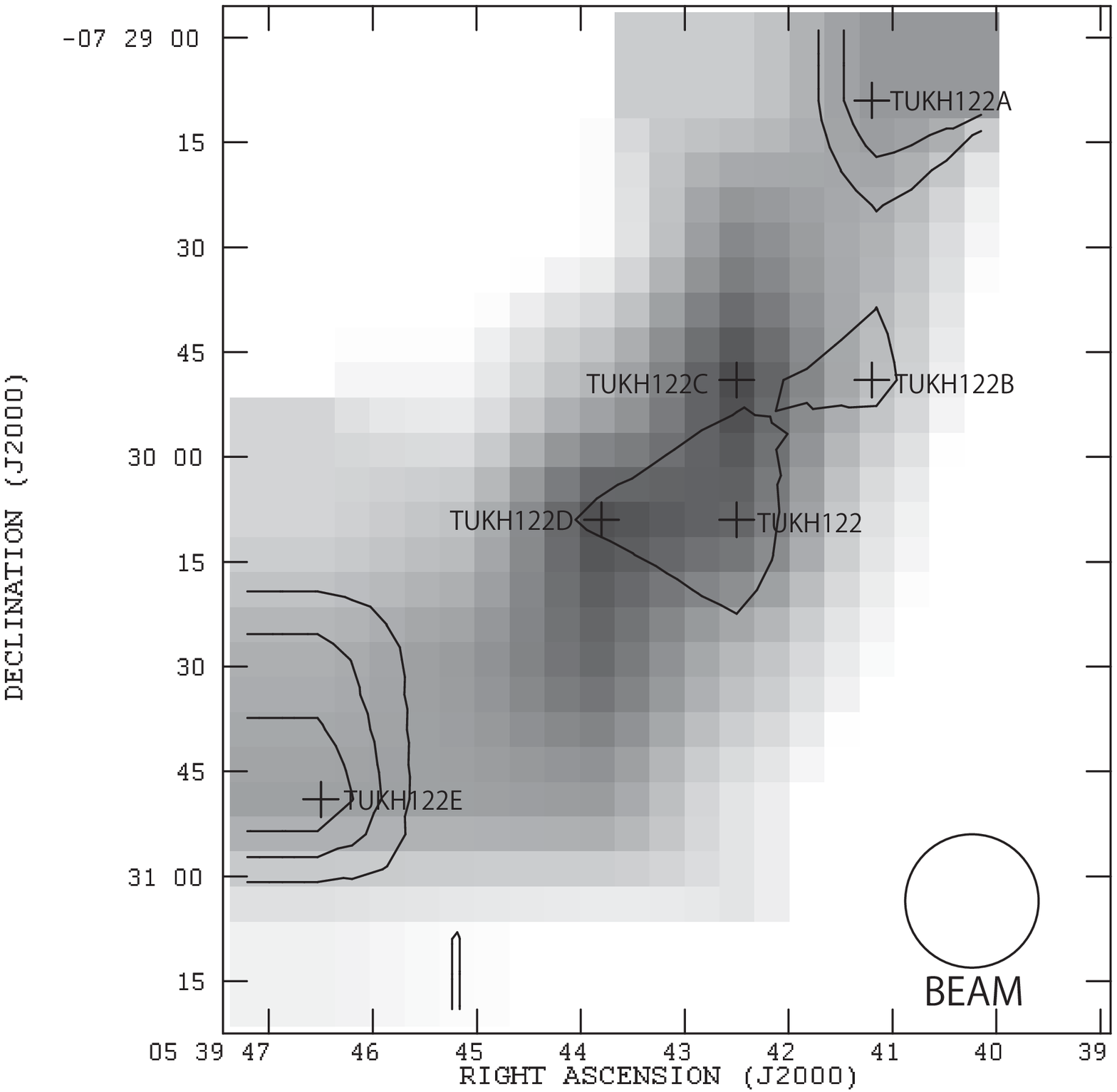}
  \end{center}
  \caption{The same as Figure 2 but for TUKH122.
The velocity range for integration is 2.5 to 5.5 km s$^{-1}$.
The 1$\sigma$ for CCS is 
2.8$\times$10$^{-2}$ K km s$^{-1}$,
and the maximum in the gray scale corresponds to 
0.88 K km s$^{-1}$.
There is no protostar.
}\label{fig:figure7}
\end{figure}

We identified 31 intensity peaks in the CCS $J_N$ = 7$_6-6_5$ and/or 
N$_2$H$^+$ $J$ = 1$-$0 $F_1$ = 2$-$1 emission
in the six molecular cloud cores.
In Table 3, we list the positions of the detected CCS $J_N$ = 7$_6-6_5$ 
intensity peaks and the line parameters of the CCS spectra measured toward these positions.
The peak intensity and linewidth are obtained through Gaussian fitting.
The rms noise level is measured for the spectrum binned by 4 channels ($\sim$ 0.4 km s$^{-1}$).
These local intensity peaks do not necessarily match the original TUKH core position.
When a local peak matches the original core position, we use the same name
(TUKH021, 097, and 122).
We add postfix "A, B, C, ..." after core number TUKH
for local intensity peaks different from the original core center position.
The NH$_3$ rotation temperature $T_{rot}$ (NH$_3$) is
taken from \citet{wil99}, and then is converted to
the gas kinetic temperature $T_k$ by using the formula of
\citet{dli03}.
The beam size used by \citet{wil99} is 43 arcsec, which is twice as large as our beam.
We assume that $T_k$
is constant over intensity peaks in each TUKH region.
\citet{dli13} obtained the distribution of $T_k$ on the basis of their NH$_3$ map toward
the OMC-2/3 region, including the TUKH003 region.
The NH$_3$ cores they identified near TUKH003A-D have $T_k$ = 14$-$16 K.
Because their NH$_3$ cores do not exactly match our CCS $J_N$ = 7$_6-6_5$ and N$_2$H$^+$ peaks 
and there is no hint of large $T_k$ variation over TUKH003A$-$D,
we simply use $T_k$ = 16 K for all of the TUKH003A$-$D peaks.
It is noted that 
NH$_3$ core
OriAN-535235-50132 at RA = 5$^h$35$^m$23$^s$.5 DEC = $-$5$\degree$01$\arcmin$32$\arcsec$
in the TUKH003 region,
which does not have CCS$J_N$ = 7$_6-6_5$ or N$_2$H$^+$ counterpart,
has $T_k = 27$ K.
Judging from $T_k$ of the six TUKH regions (see Table 3),
cores TUKH097 and 117 (and also probably 122) show similar 
conditions to those found in dark cloud cores.
Cores TUKH003, 088, and 021 are warmer than dark cloud cores
(with $T_k$ increasing from 16 to 30 K in this order).

We carry out the large velocity gradient (LVG) model
by using the RADEX software \citep{van07}
to characterize the CCS $J_N$ = 7$_6-6_5$ detected cores.
The collision rates for CCS are taken from
\citet{wol97}.
We detected 
CCS $J_N$ = 7$_6-6_5$ at 81.505208 GHz
in 7 cores out of the 14 cores toward which 
CCS $J_N$ = 4$_3-3_2$ at 45.379033 GHz
was detected.
From the LVG analysis, the cores which were not detected in 
CCS $J_N$ = 7$_6-6_5$ are found to have lower densities
($n <$ 1$\times$10$^5$ cm$^{-3}$) than the CCS $J_N$ = 7$_6-6_5$ detected cores
($n \gtrsim$ 1$\times$10$^5$ cm$^{-3}$).
In the LVG calculations, we 
assume that the CCS $J_N$ = 4$_3-3_2$ and CCS $J_N$ = 7$_6-6_5$ 
line emission has similar spatial 
extent and distribution within the single-dish beam of the observations,
and we do not correct for the difference in the beam size.
According to \citet{tat08}, the average density of N$_2$H$^+$ $J$ = 1$-$0 cores
is of order 1$\times$10$^5$ cm$^{-3}$.
Then, the CCS $J_N$ = 7$_6-6_5$ and N$_2$H$^+$ $J$ = 1$-$0 emission seems to trace
similar density regions.

In TUKH 088, 097, and 117, there is a trend for the CCS emission to be stronger 
where the protostar does not exist,
while the N$_2$H$^+$ $J$ = 1$-$0 $F_1$ = 2$-$1 emission tends to be stronger 
toward the protostar (candidate).
These tendencies are similar to those found in dark clouds \citep{ben98}.
In some cases, the CCS emission is detected toward the protostar as well as
the N$_2$H$^+$ emission.
Note that the N$_2$H$^+$ emission is much stronger than the CCS $J_N$ = 7$_6-6_5$ emission in general.
A possible explanation for the CCS $J_N$ = 7$_6-6_5$ emission toward star-forming peaks 
is that CCS also exists in evolved
molecular gas as a secondary late-stage peak due to CO depletion
in the chemical evolution
(\cite{zli02} and references therein; \cite{lee03}).

In the TUKH003 and TUKH021 regions,
there are protostars which do not accompany
the N$_2$H$^+$ $J$ = 1$-$0 $F_1$ = 2$-$1 emission.
These are Megeath 2017, 2044, 2050, 2060, 2073, 2091, 2427, 2433, 2437,
and 2451. The reason will be investigated in the subsequent subsection.

To see the evolutionary stage of the intensity peaks, 
we check the association with protostars.
We assume that starless intensity peaks are young, while
star-forming peaks are more evolved.
As a criterion, we classify 
CCS $J_N$ = 7$_6-6_5$ and/or N$_2$H$^+$ $J$ = 1$-$0 $F_1$ = 2$-$1 intensity peaks
as ``starless'' and ``star-forming''
when the distance between the position of
the protostar and the core intensity peak
is 
$>$ 30 arcsec and $\leq$ 30 arcsec, respectively.
We explain why we adopt this criterion.
According to \citet{tat08} the average radius (half of FWHM) of 
N$_2$H$^+$ $J$ = 1$-$0 cores is 39$\pm$ 12 arcsec.
This provides the size scale of the embedded region.
Next, we consider the proper motion of the protostar with respect to
the parent molecular cloud core after its birth.
We take the age of the Class I protostar to be of order
1$-$3$\times$10$^5$ yr (e.g., \cite{gre94}).
We assume that the one-dimensional velocity dispersion (1$\sigma$) of the protostar
with respect to the core center is equal to that of the velocity 
dispersion of the linewidth of the emission toward the core
(FWHM/(2 $\sqrt{2 \ln 2}$)).
We take an FWHM linewidth of 0.5 km s$^{-1}$ as a representative number
(The FWHM linewidth of N$_2$H$^+$ $J$ = 1$-$0 for all the cores except TUKH021
is 0.63$\pm$0.23 km s$^{-1}$ as seen from Table 4; to be explained later).
Then,
the maximum shift perpendicular to the line of sight of the 
Class I protostar with respect to the core center is
0.07 pc (30 arcsec).
Finally, we need take into account the spatial resolution
of our observations. Our observations were carried out on a 20 arcsec grid
with 19.1$\pm$0.3 arcsec beam, and this will determine the positional accuracies 
of the intensity peaks.
Taking these into account, 
we adopt the criterion of 30 arcsec (0.07 pc) for association with 
the Spitzer source.

\subsection{Individual Regions} 

In the TUKH003 region (Figure 2),
three Spitzer sources are located near N$_2$H$^+$ $J$ = 1$-$0 $F_1$ = 2$-$1 emission.
TUKH003B is associated with 
Megeath 2442 and 2446, which correspond
to X-ray emitting protostars
TKH10 and TKH8 discovered by
\citet{tsu01} (see also \cite{tsu04}), respectively.
\citet{tat08} observed 
HC$_3$N $J$ = 5$-$4 with a 38 arcsec beam and N$_2$H$^+$ $J$ = 1$-$0 
with a 18 arcsec beam, and found a
displacement between the molecular emission from 
these molecules
over 2 arcmin (0.3 pc) scale in this region. 
The present observations with a 19 arcsec beam show a
more complicated distribution.
Megeath 2440 is also close to TUKH003B.
On the other hand, protostars Megeath 2427, 2433, 2437, and 2451 are not close to
any bright CCS $J_N$ = 7$_6-6_5$ or N$_2$H$^+$ $J$ = 1$-$0 $F_1$ = 2$-$1 local peaks  
(i.e., not within a radius of $\leq$ 30 arcsec). 
Figure 8 compares the CCS $J_N$ = 7$_6-6_5$ distribution with the dust continuum emission
obtained by
\citet{joh99} with 
the James Clerk Maxwell Telescope (JCMT) and the SCUBA 
(Submillimeter Common-User Bolometer Array).
The dust continuum map was obtained with a 14 arcsec beam, which is close
to the resolution of our observations.
Structures with scales larger than 65 arcsec are missing in their continuum map
due to the observing method.
The brightest dust continuum source corresponds to MMS6 \citep{chi97},
CSO10 \citep{lis98}, and OriAN-535235-50132 \citep{dli13}. 
Except for this brightest source, the dust continuum distribution is
very similar to that of N$_2$H$^+$ $J$ = 1$-$0 $F_1$ = 2$-$1.
The N$_2$H$^+$ $J$ = 1$-$0 $F_1$ = 2$-$1 distribution traces 
relatively well the column density of the high density region
affected very little by protostellar feedback,
as reported toward 
low-mass pre-stellar cores \citep{cas99}.
However, the brightest dust continuum source OriAN-535235-50132 is associated with molecular gas with $T_k = 27$ K \citep{dli13}, higher than the CO 
evaporation temperature of CO from the mantles of dust grains 
at about 25K \citep{col04}.
Therefore, CO evaporated from grain surfaces destroys N$_2$H$^+$ significantly 
in warm regions with $T_{dust} >$ 25 K \citep{lee04}. 
Then, it is expected that the N$_2$H$^+$ emission
is not prominent in warmer regions where stars have already formed.
This could be a reason why
protostars Megeath 2427, 2433, and 2437 are not 
associated with N$_2$H$^+$ $J$ = 1$-$0 $F_1$ = 2$-$1 local peaks (see Figure 2).

\begin{figure}
  \begin{center}

    \FigureFile(150mm,150mm){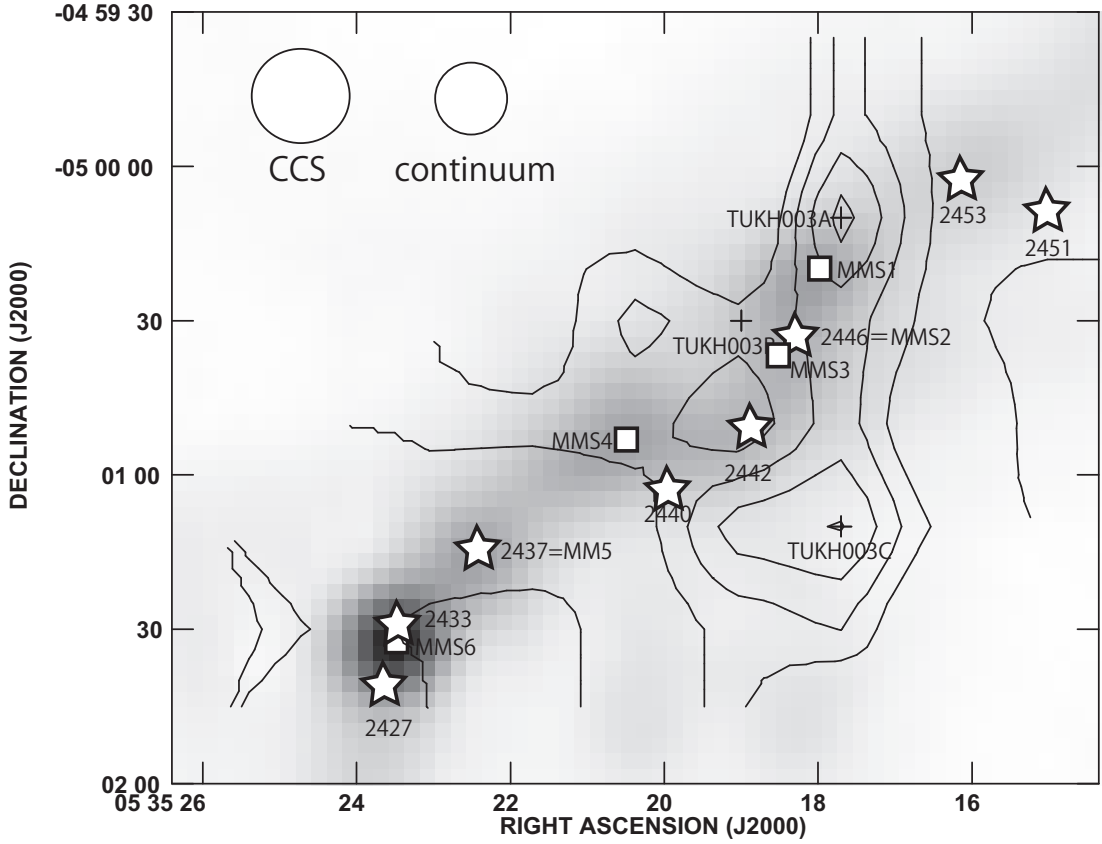}

  \end{center}
  \caption{
The CCS $J_N$ = 7$_6-6_5$ velocity-integrated intensity map
is superimposed
on the gray-scale map of the 850 $\mu$m dust continuum emission
of \citet{joh99} for
TUKH003.
The dust continuum map was obtained with a 14 arcsec beam, which is close to the angular resolution of our observations (19 arcsec).
Structures with scales larger than 65 arcsec are 
missing in the continuum map due to the observing technique used (see \cite{joh99})
The maximum in the gray scale corresponds to 
7.4 Jy beam$^{-1}$.
The open box signs represent the 1.3 mm dust continuum sources cataloged by \citet{chi97} (source names with the prefix MMS).
When the 1.3 mm dust continuum coincides with the protostars, the open box signs are omitted and
only the source names are shown.
}\label{fig:figure8}
\end{figure}

The TUKH021 region (Figure 3) is located
a few arcmin north of Orion KL.
There is some correlation between the distribution
of CCS $J_N$ = 7$_6-6_5$ and N$_2$H$^+$.
Megeath 2037, 2069, and 2106 are located near the N$_2$H$^+$ $J$ = 1$-$0 $F_1$ = 2$-$1 emitting regions.
On the other hand, protostars Megeath 2017, 2044, 2050, 2073, and 2091 
are not close to
any of the N$_2$H$^+$ $J$ = 1$-$0 $F_1$ = 2$-$1 local peaks.
Megeath 2060 is close to TUKH021B, which is a CCS $J_N$ = 7$_6-6_5$ peak.
It is not clear whether this is chance coincidence. 
Figure 9 compares the CCS $J_N$ = 7$_6-6_5$ distribution with the dust continuum emission
by \citet{joh99}.
Megeath 1971, 2044, and 2060 are located along the south-east dust ridge 
seen in this figure.
Figure 10 compares the CCS $J_N$ = 7$_6-6_5$ distribution with the C$^{18}$O $J$ = 1$-$0 map
carried out with the Nobeyama 45 m
radio telescope on a 17 arcsec grid with a 15 arcsec beam
\footnote{The data available at
http://alma.mtk.nao.ac.jp/$\sim$kt/fits.html}.
The grid spacing and beam size of the C$^{18}$O observations are close to 
those used in the present observations
(20 and 19.1 arcsec, respectively).
In general, the dust continuum and  C$^{18}$O distribution is 
more or less similar to the
N$_2$H$^+$ distribution, but shows appreciable
differences from the CCS $J_N$ = 7$_6-6_5$ distribution.
Warm dust temperature could be a reason why
protostars Megeath 2044, 2060, and 2091 
are not 
associated with N$_2$H$^+$ $J$ = 1$-$0 $F_1$ = 2$-$1 local peaks.
We adopted the gas kinetic temperature of $T_k$ = 30 K
obtained toward original core center position TUKH021 (and local intensity peak
position) with a 43 arcsec beam, for the overall TUKH021 region.
Judging from Figure 8 of \citet{wis98}, whose spatial resolution 
is 8.5$\times$9.0 arcsec, the gas kinetic temperature $T_k$
changes in a complicated way from 18 to 45 K between 
DEC(J2000.0) = 
$-$05$\degree$20$\arcmin$00$\arcsec$ 
and
$-$05$\degree$19$\arcmin$00$\arcsec$.
It seems that this region has small-scale 
gas kinetic temperature variations. Since we do not have
gas kinetic temperature measurements which exactly match
our beam and sampling, it is hard to assign the gas kinetic temperature to each 
local intensity peak.
The excitation temperature of N$_2$H$^+$ toward TUKH021 is 
20.8$\pm$1.8 K.  Then, the gas kinetic temperature $T_k$ toward TUKH021 
with a 19 arcsec beam will be $>$ 21 K.

\begin{figure}
  \begin{center}

    \FigureFile(150mm,150mm){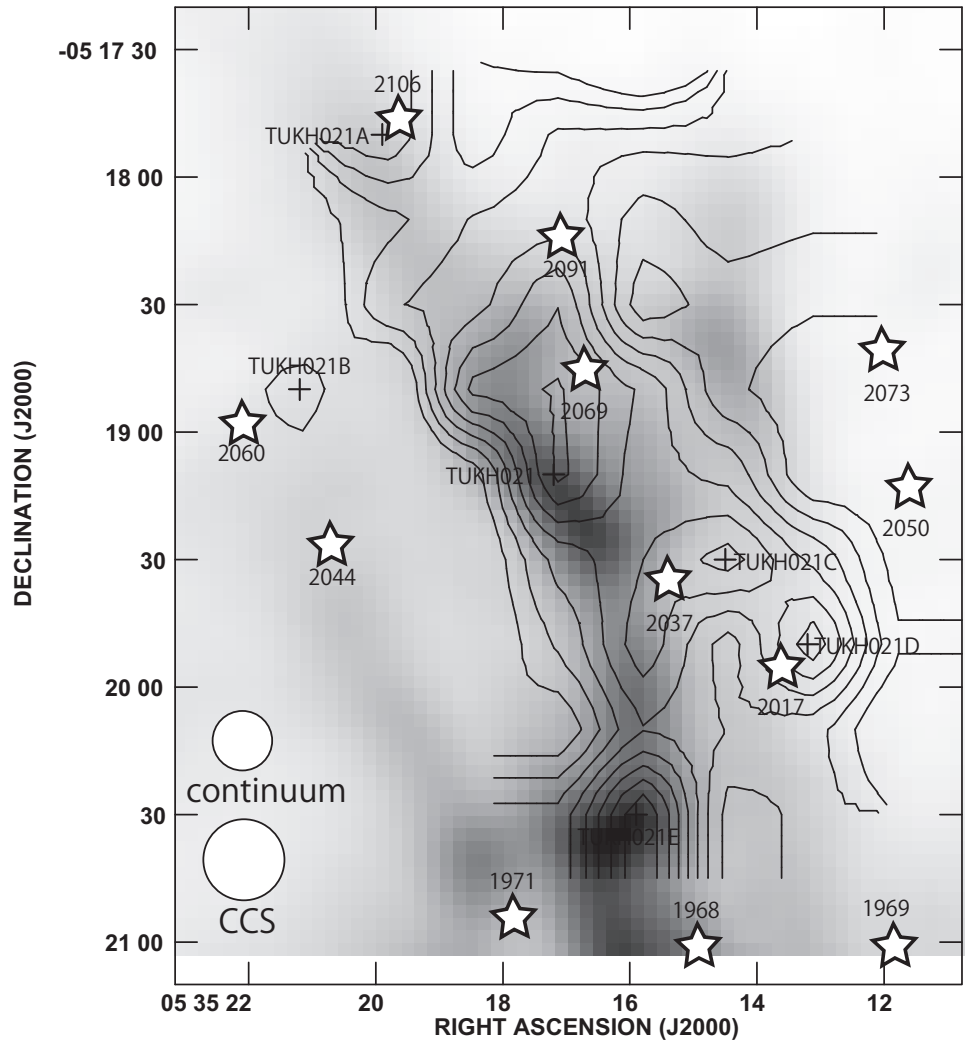}

  \end{center}
  \caption{
The same as Figure 8 but for
TUKH021.
The maximum in the gray scale corresponds to 
11.7 Jy beam$^{-1}$.
}\label{fig:figure9}
\end{figure}

\begin{figure}
  \begin{center}

    \FigureFile(150mm,150mm){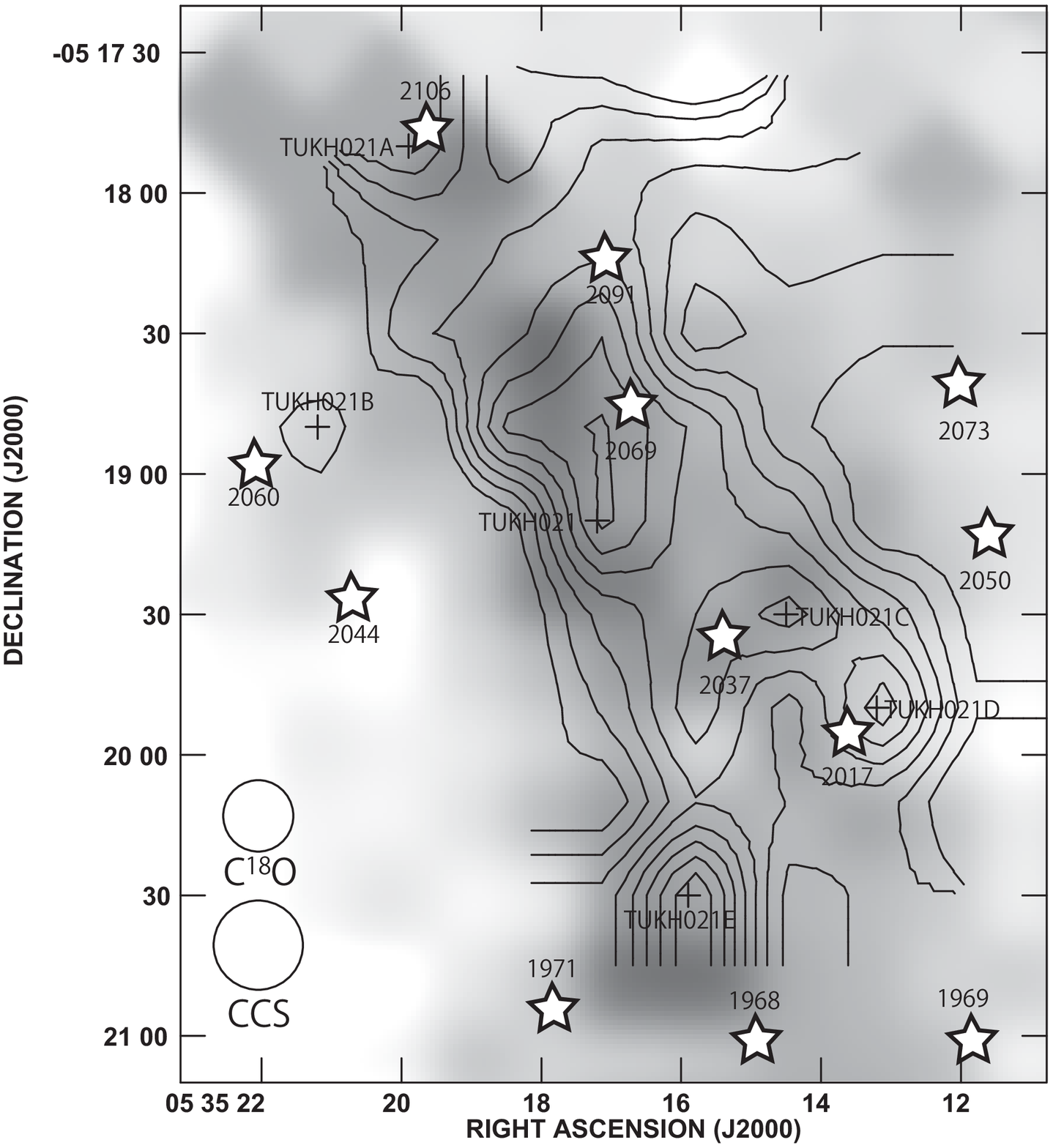}

  \end{center}
  \caption{
The CCS $J_N$ = 7$_6-6_5$ velocity-integrated intensity map
is superimposed
on the gray-scale map of C$^{18}$O $J$ =1$-$0
for
TUKH021.
The C$^{18}$O data were obtained on a 17 arcsec grid with a 15 arcsec beam.
The velocity range for C$^{18}$O is 5.0 to 13.7 km s$^{-1}$.
The maximum in the gray scale corresponds to 
5.1 K km s$^{-1}$.
}\label{fig:figure10}
\end{figure}

In the TUKH088 region (Figure 4),
faint candidate protostar Megeath 826 is associated with TUKH088B, 
where both the N$_2$H$^+$ $J$ = 1$-$0 $F_1$ = 2$-$1 and CCS  $J_N$ = 7$_6-6_5$ intensity peaks are detected.
The CCS $J_N$ = 7$_6-6_5$ emission is extended toward the south, where no protostar is found.
In the north, there is one CCS $J_N$ = 7$_6-6_5$ intensity peak without a 
protostar.

In the TUKH097 region (Figure 5),
two protostars, Megeath 638 and 640, are associated with 
local peaks of the N$_2$H$^+$ $J$ = 1$-$0 $F_1$ = 2$-$1 emission,
TUKH097, TUKH097A, and TUKH097B,
when we adopt the 30 arcsec criterion.
It seems the CCS $J_N$ = 7$_6-6_5$ emission is also peaked near these two protostars.
We found two other CCS intensity peaks, TUKH097D and TUKH097E
(for TUKH097E, 
contours are not closed due to the limited observed area), 
which are not associated with protostars.

The TUKH117 region (Figure 6) is associated with Haro 4-255 FIR, 
which drives a molecular 
outflow \citep{eva86,lev88}, and with 
the VLA point source Haro 4-255 VLA1 \citep{ang92}.
It is likely that Megeath 545 and 551 correspond to Haro 4-255 VLA1 and Haro 4-255 FIR,
respectively.
The N$_2$H$^+$  $J$ = 1$-$0 $F_1$ = 2$-$1 emission is distributed around these two Spitzer sources.
It is found that the CCS $J_N$ = 7$_6-6_5$ and N$_2$H$^+$ distribution are anticorrelated.
This is consistent with the tendency found in dark clouds \citep{aik01}.
The CCS $J_N$ = 7$_6-6_5$ and N$_2$H$^+$  $J$ = 1$-$0 $F_1$ = 2$-$1 distribution in the present observation
roughly correspond to the C$^{18}$O and NH$_3$ distribution in low-resolution maps
summarized in \citet{tat93b}, respectively.

The TUKH122 region (Figure 7) is starless.
In \citet{tat10}, TUKH122 is the most CCS $J_N$ = 4$_3-3_2$ intense core,
although the CCS column density is close to the average 
obtained from the CCS $J_N$ = 4$_3-3_2$ detected cores.
The N$_2$H$^+$  $J$ = 1$-$0 $F_1$ = 2$-$1 and CCS $J_N$ = 7$_6-6_5$ intensity peaks are located within 1 arcmin
region near the map center.  
It is most likely that this 1 arcmin region represents
a physical density peak, because these molecules, which sometimes show
different distribution suggesting the chemical evolution,
coexist in the 1 arcmin region.
The LTE mass and virial mass of the core are estimated to be 49 and 26 M$_{\odot}$, respectively, from CS $J$ = 1$-$0 observations \citep{tat93a}.
Because we detected the emission from the late-type molecule N$_2$H$^+$ and there is no protostar,
it is possible that the TUKH122 region is on the verge of star formation (cf. \cite{cas99}).

\subsection{Hyperfine Line Fitting, Integrated Intensity Ratios, and Column Density Ratios}

We fit the hyperfine component model to the N$_2$H$^+$  $J$ = 1$-$0 spectrum,
and derive the optical depth, LSR velocity, linewidth, and excitation temperature.
The intrinsic line strength of the hyperfine components is adopted from \citet{tin00}.
The results are shown in Table 4.
The optical depth $\tau_{TOT}$ is the sum of the optical depths of all the hyperfine components.
The velocity-integrated intensity $W$ = $\int T_A^* dv$ of 
the main
hyperfine component group
N$_2$H$^+$ $J$ = 1$-$0 $F_1$ = 2$-$1
is also listed.
Blank cells represent that the hyperfine line fitting is not successful
or that derived parameters have too large fitting
errors. 
Figures 11 to 17 show some examples with the results of hyperfine line fitting method.
In some cases, the hyperfine line fitting is not very good, because regions contain
two velocity components or skewed velocity profiles.  
We have also tried two velocity component fitting, 
which provides us with better fitting.
On the other hand, two velocity component Gaussian fitting to the CCS spectrum is hard to
carry out because of lower signal-to-noise ratios in the CCS spectrum.
There is no guarantee that the CCS emission has the same velocity components as the N$_2$H$^+$
emission.
Because the main purpose of the present study is a comparison of the CCS and N$_2$H$^+$ emission,
we adopt the single-velocity hyperfine line fitting in this paper.
The N$_2$H$^+$ LSR velocity from single-velocity fitting is more consistent with
the CCS LSR velocity than the N$_2$H$^+$ LSR velocities from two-velocity fitting.
For example, TUKH003B has an N$_2$H$^+$ and CCS velocity of 11.05 and 10.98 km s$^{-1}$, respectively, while the two-velocity fitting for N$_2$H$^+$ gives 10.58 and 11.18 km s$^{-1}$.
In our future, separate paper, we plan to investigate the core dynamics on the basis of the N$_2$H$^+$
spectra and the result of multi-velocity-component hyperfine line fitting will be presented there.

\begin{figure}
  \begin{center}
    \FigureFile(150mm,150mm){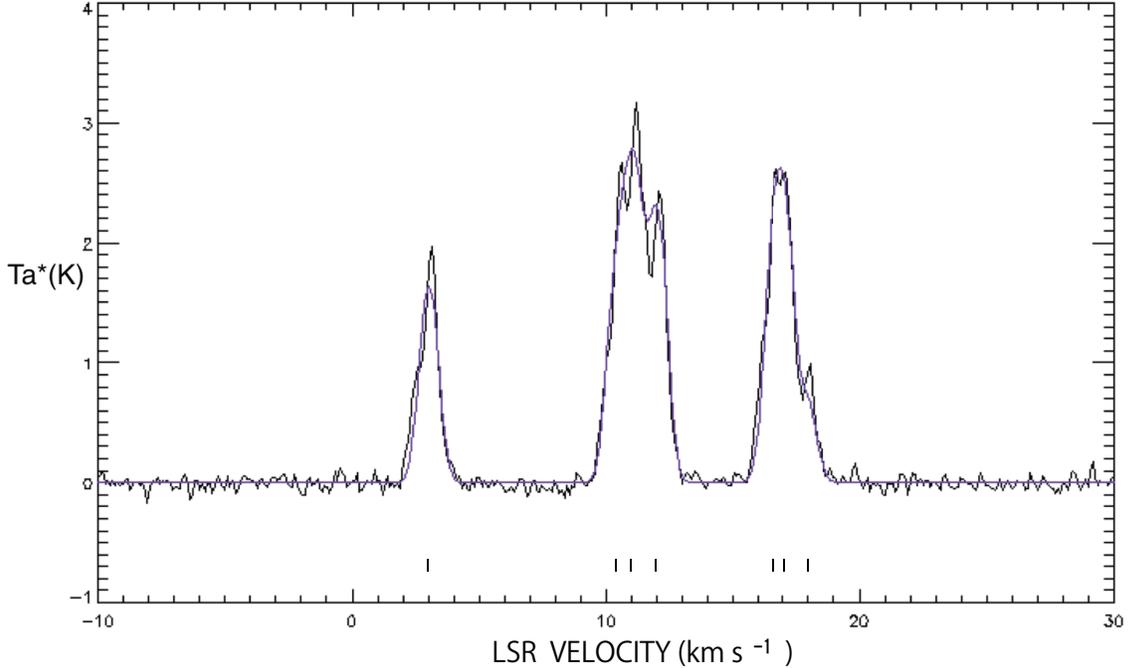}
  \end{center}
  \caption{Hyperfine line fitting result for the N$_2$H$^+$  $J$ = 1$-$0 spectrum
for TUKH003B.
Seven short vertical bars are shown to illustrate the velocity offset
corresponding to the frequency offset
of the hyperfine components.
}\label{fig:figure11}
\end{figure}

\begin{figure}
  \begin{center}
    \FigureFile(150mm,150mm){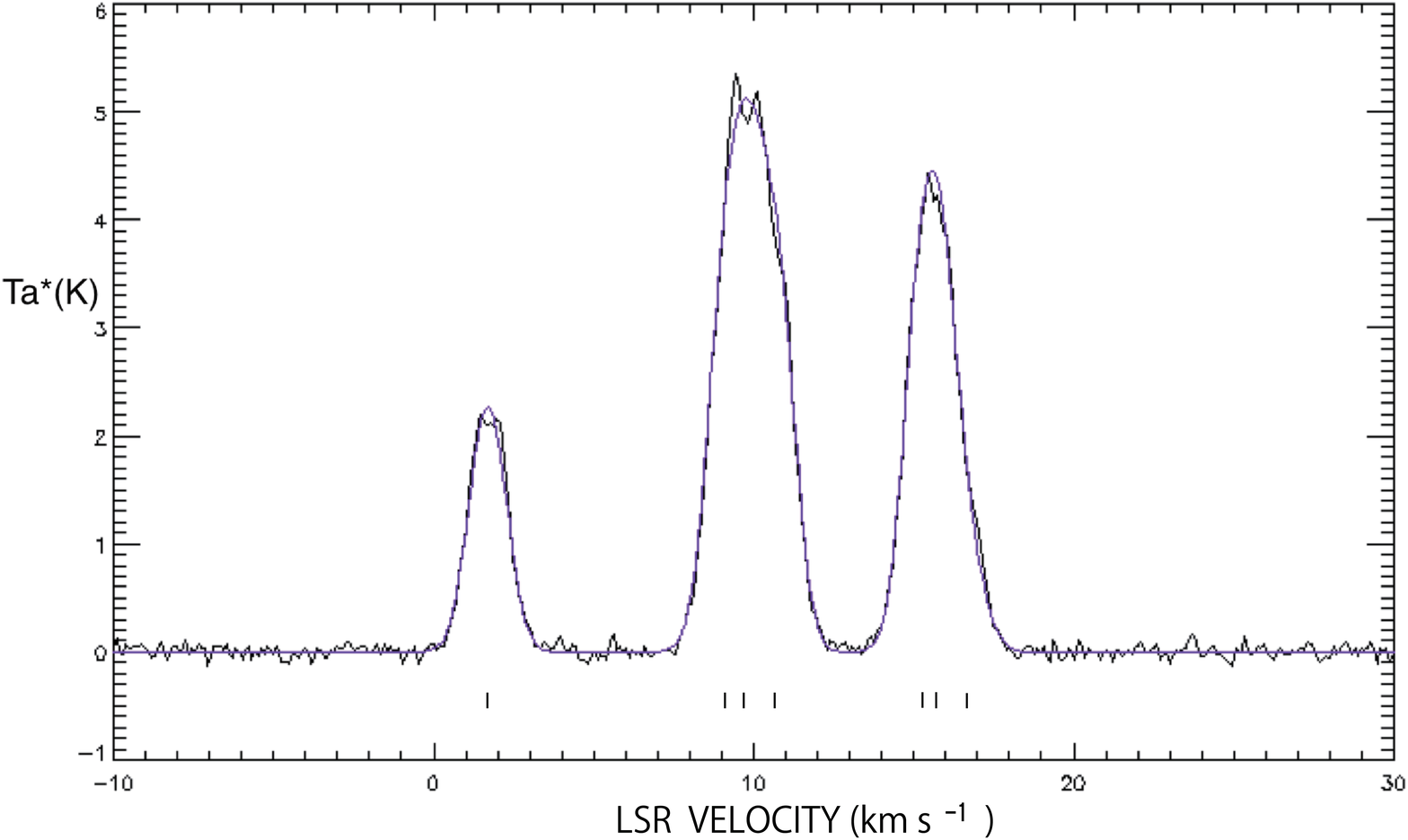}
  \end{center}
  \caption{The same as Figure 11 but
for TUKH021. 
}\label{fig:figure12}
\end{figure}

\begin{figure}
  \begin{center}
    \FigureFile(150mm,150mm){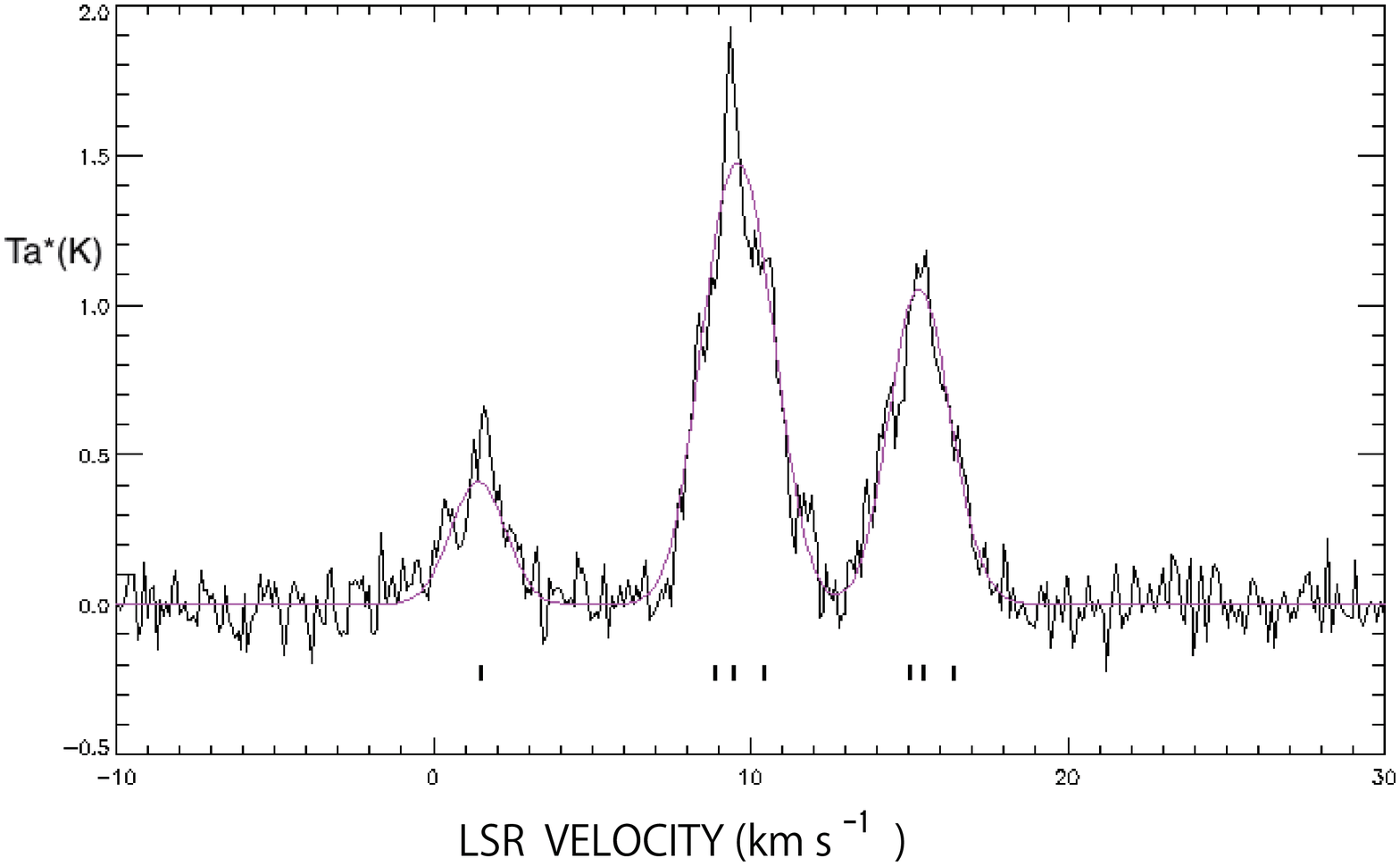}
  \end{center}
  \caption{The same as Figure 11 but for TUKH021C.
}\label{fig:figure13}
\end{figure}

\begin{figure}
  \begin{center}
    \FigureFile(150mm,150mm){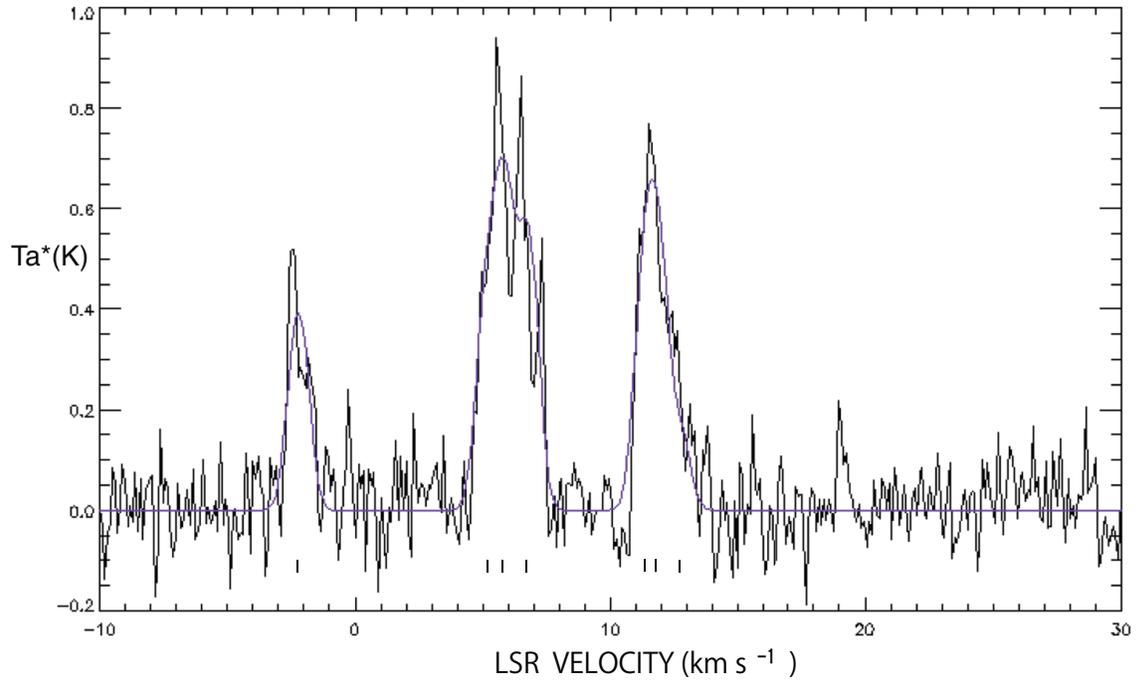}
  \end{center}
  \caption{The same as Figure 11 but for TUKH088B.
}\label{fig:figure14}
\end{figure}

\begin{figure}
  \begin{center}
    \FigureFile(150mm,150mm){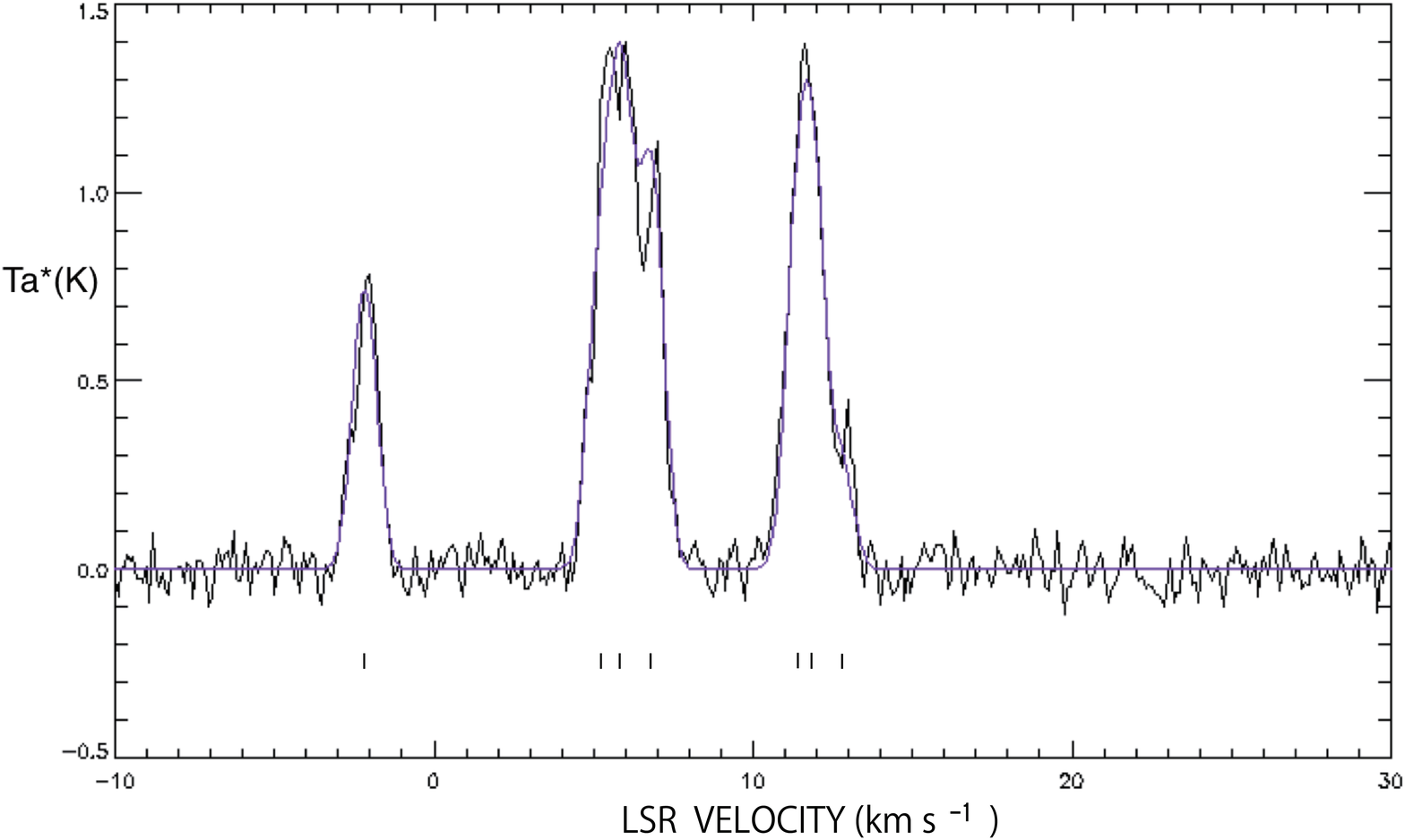}
  \end{center}
  \caption{The same as Figure 11 but for TUKH097C.
}\label{fig:figure15}
\end{figure}

\begin{figure}
  \begin{center}
    \FigureFile(150mm,150mm){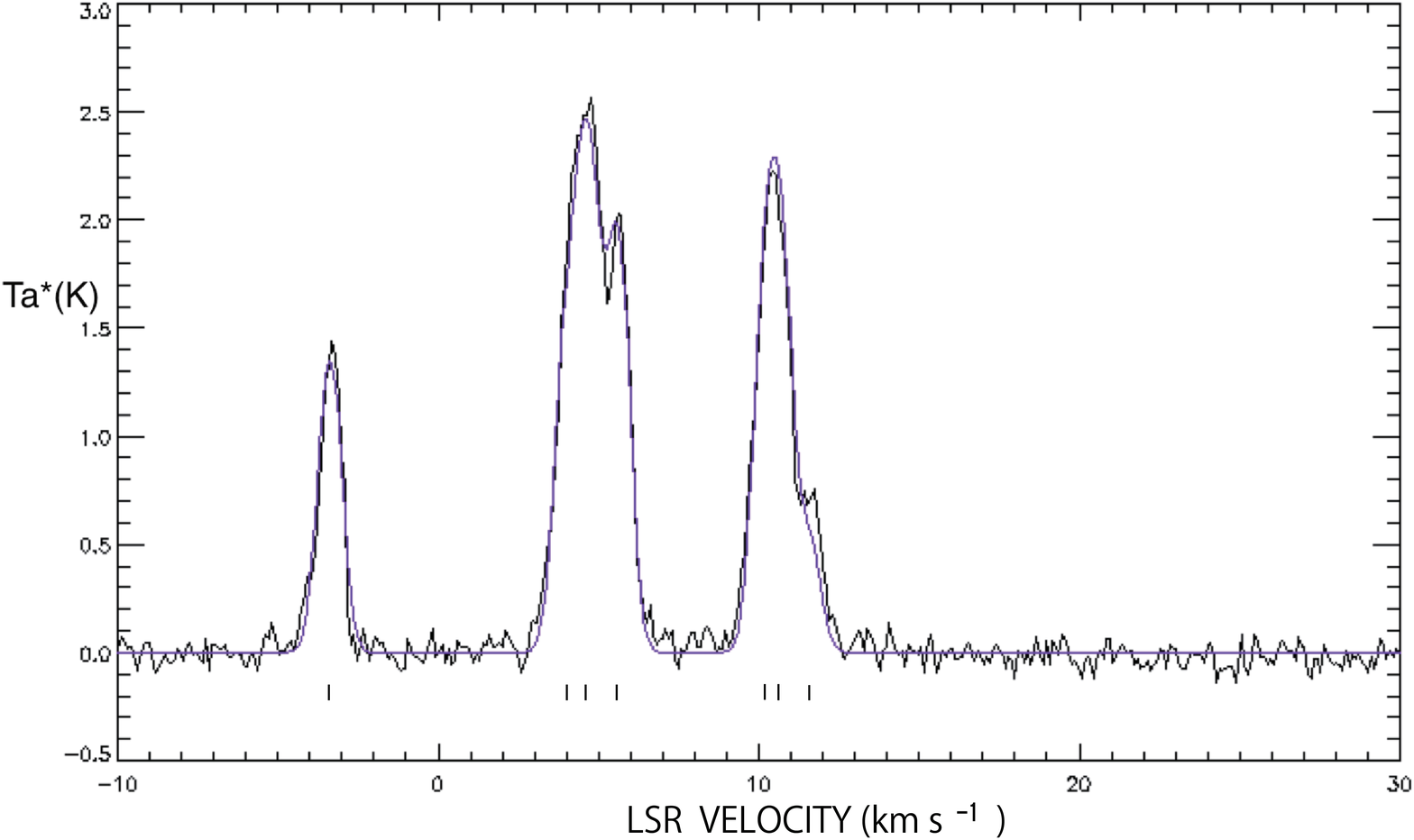}
  \end{center}
  \caption{The same as Figure 11 but for TUKH117B.
}\label{fig:figure16}
\end{figure}

\begin{figure}
  \begin{center}
    \FigureFile(150mm,150mm){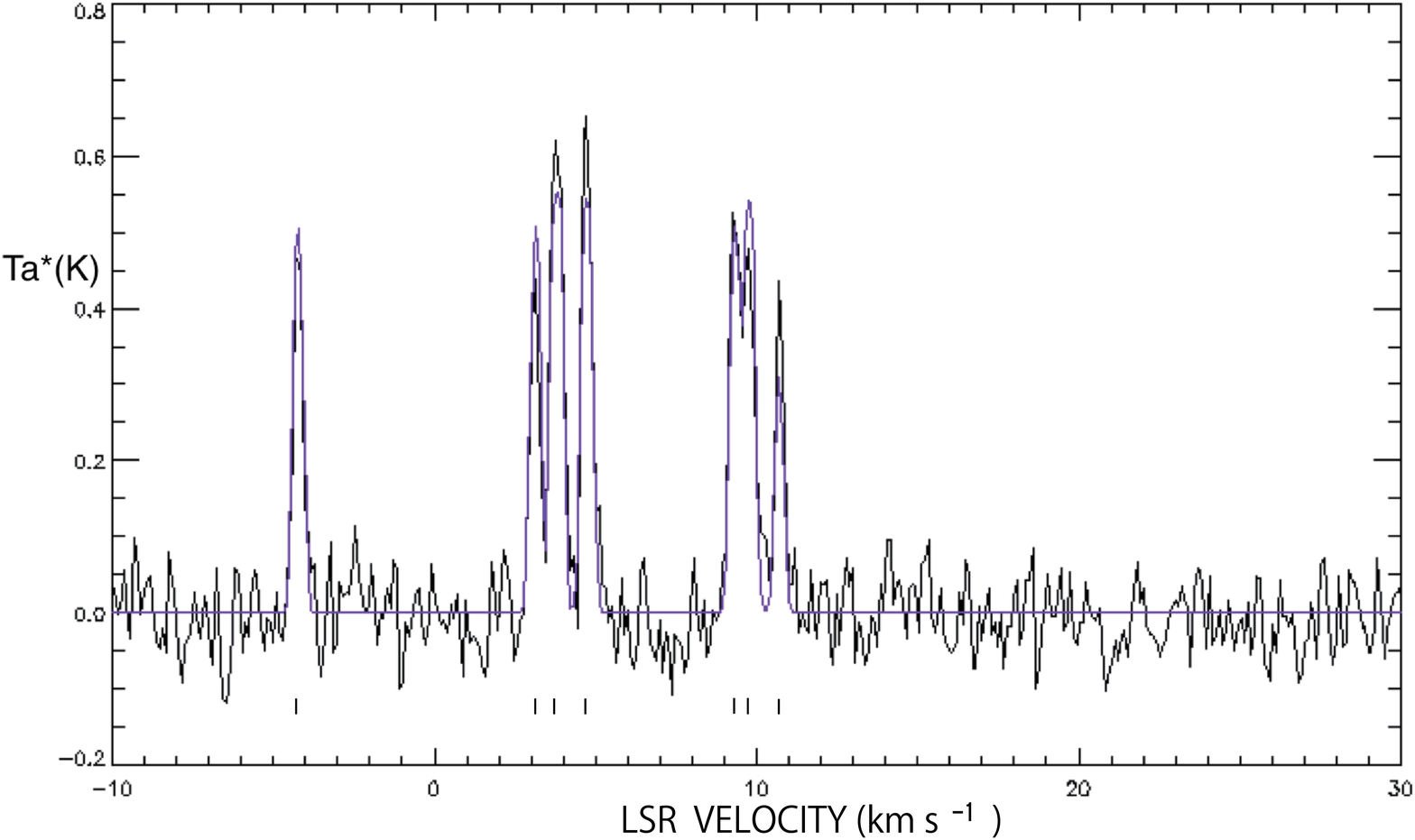}
  \end{center}
  \caption{The same as Figure 11 but for TUKH122.
}\label{fig:figure17}
\end{figure}

Figure 18 compares 
the excitation temperature $T_{ex}$ 
derived from the N$_2$H$^+$  $J$ = 1$-$0  hyperfine line fitting method
against $T_k$
derived from NH$_3$ \citep{wil99}.
The open circle and filled circle represent star forming and starless peaks,
respectively.
When the emission is more optically thin, we cannot constrain the optical depth well 
and the error bars are large (one core at $T_k$ = 29 K has $\tau_{TOT}$ = 0.7$\pm$0.5 and $\tau$ for each component is 0.1$-$0.2).
In general, $T_{ex}$ (N$_2$H$^+$) is appreciably lower than $T_k$.
We conclude that the N$_2$H$^+$  $J$ = 1$-$0 levels are only subthermally excited.

\begin{figure}
  \begin{center}
    \FigureFile(150mm,150mm){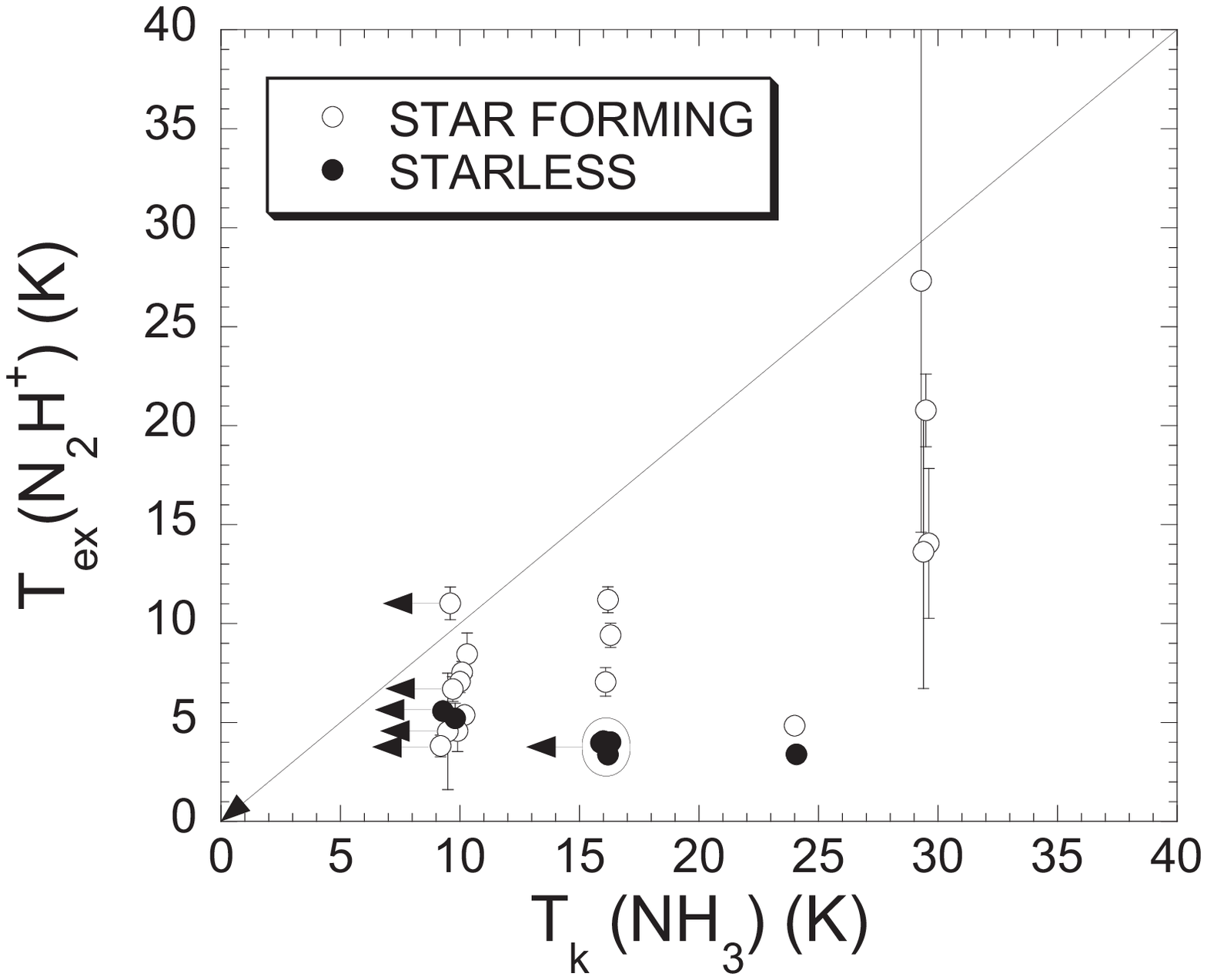}
  \end{center}
  \caption{The N$_2$H$^+$  $J$ = 1$-$0 excitation temperature is
plotted against the gas kinetic temperature $T_k$.
The gas kinetic temperature is obtained from the NH$_3$ rotation temperature in
\citet{wil99} with the conversion formula of \citet{dli03}.
The horizontal arrow represents the upper limit to $T_k$.
The vertical error bar represents the 1$\sigma$ error corresponding to the line optical depth 
in the N$_2$H$^+$  $J$ = 1$-$0 hyperfine line fitting.
Circles are slightly shifted so that error bars do not overlap with each other.
The straight line $T_{ex} = T_k$ is shown.
}\label{fig:figure18}
\end{figure}

In Figure 19, the 
ratio of the velocity-integrated intensity
of the N$_2$H$^+$ main hyperfine group $J$ = 1$-$0 $F_1$ = 2$-$1 to that of the CCS $J_N$ = 7$_6-6_5$ emission is
plotted against $T_k$.
It seems that star forming peaks tend to have larger ratios.

\begin{figure}
  \begin{center}
    \FigureFile(150mm,150mm){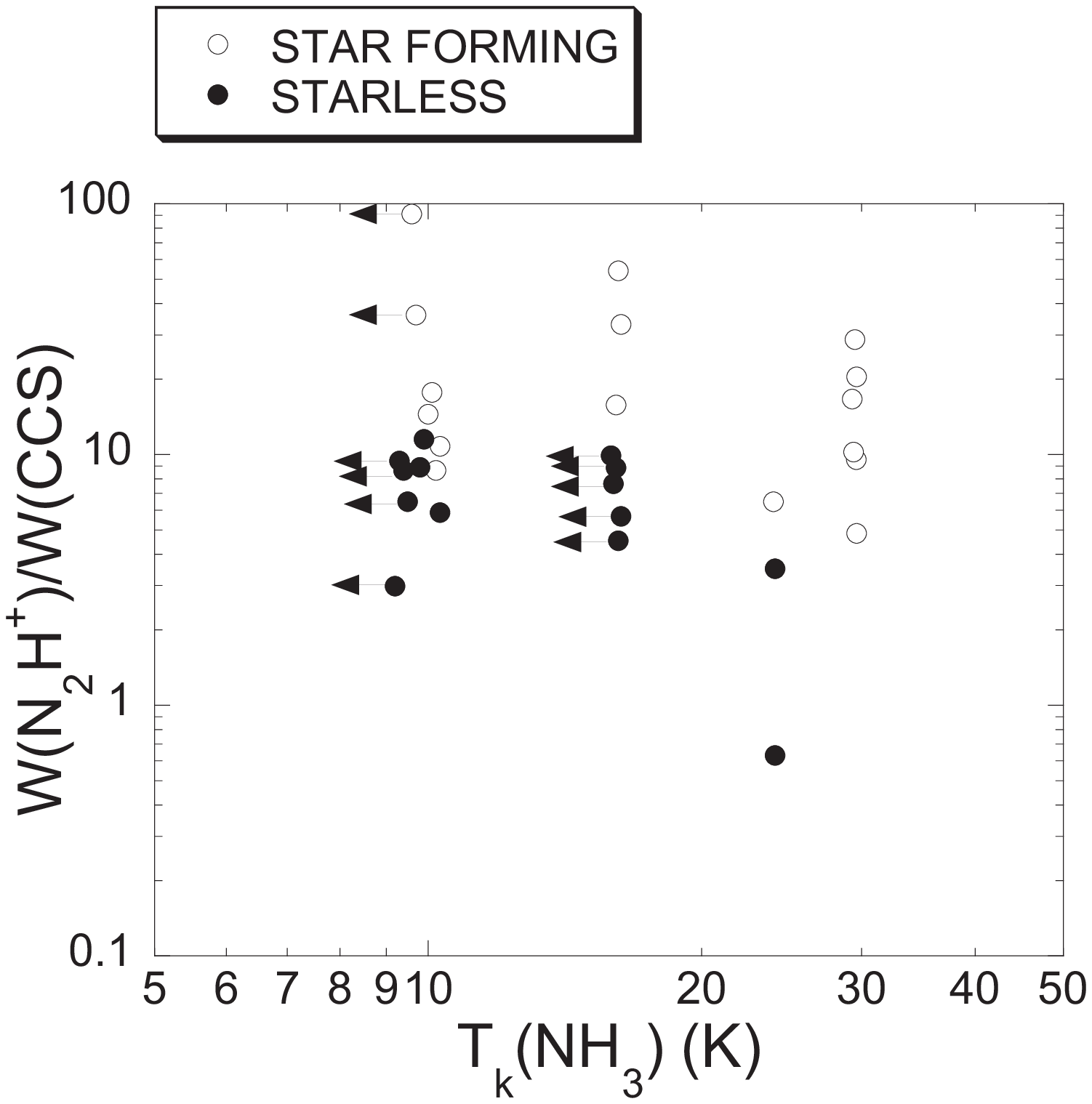}
  \end{center}
  \caption{The ratio of the velocity-integrated intensity
of N$_2$H$^+$ main hyperfine group $J$ = 1$-$0 $F_1$ = 2$-$1 to that of CCS is
plotted against the gas kinetic temperature $T_k$. 
}\label{fig:figure19}
\end{figure}

Next, we compare the column density of 
N$_2$H$^+$ with that of CCS.
The column density is calculated by assuming
local thermodynamic equilibrium (LTE). The formulation
can be found, for example, in
\citet{suz92}.
We use the N$_2$H$^+$ line parameters derived from the hyperfine line fitting
method.
The excitation temperature $T_{ex}$ (CCS) is assumed to be equal to
that for N$_2$H$^+$ when the hyperfine line fitting to the N$_2$H$^+$
$J$ = 1$-$0 spectrum
at the same position is successful.
In some cases, LSR velocities of the 
N$_2$H$^+$ $J$ = 1$-$0 and CCS $J_N$ = 7$_6-6_5$ emission
at the same position differ slightly, which means that emitting regions
are different.  Even in such cases, we simply assume the same excitation 
temperature as a best guess.
When the hyperfine line fitting is not successful due to low signal to noise ratios,
we adopt $T_{ex}$ (CCS) = $T_k$ (NH$_3$)/2 (when \cite{wil99} 
provide the upper limit
to $T_{rot}$,
we simply adopt this upper limit value).
In one case, the error bar is large.
This comes from a large error in the N$_2$H$^+$ optical depth because
the optical depth is thin (one core at $T_k$ = 29 K has $\tau_{TOT}$ = 0.7$\pm$0.5 and $\tau$ for each component is 0.1$-$0.2).
Table 5 summarizes the derived column densities.
Blank cells indicate that column densities could not be determined 
due to the low signal-to-noise ratio in the N$_2$H$^+$ $J$ = 1$-$0
spectra.
Figure 20 shows the 
ratio of the column density
of N$_2$H$^+$ 
to that of CCS 
against $T_k$.
The number of points in Figure 20 is smaller than that in Figure 19,
because in some cases hyperfine line fitting is not successful and we cannot obtain
the N$_2$H$^+$ column density.
One circle at $T_k$ = 29 K has a very large error bar, because the spectrum is close to optically
thin limit and the optical depth cannot be constrained precisely.
It is found
that 
the column density ratio of $N$(N$_2$H$^+$)/$N$(CCS) is
low toward starless peaks while it is high
toward star-forming peaks.
This is very similar to the tendency found in dark clouds 
($T_k$ $\sim$ 10 K) \citet{ben98}.
The criterion found in the Orion A GMC is $N$(N$_2$H$^+$)/$N$(CCS) $\sim 2-3$.
That is, 
$N$(N$_2$H$^+$)/$N$(CCS) is $\lesssim 2-3$ in starless peaks, and
$\gtrsim 2-3$ in star-forming peaks.
TUKH097E, 117F, 122C, 122, and 122D have larger column density ratios
among starless peaks, and their values are close to the criterion of 
$2-3$ between starless
and star-forming peaks found in the present study.
It is possible that these represent the sites for near future star formation.
TUKH122C, 122, and 122D are of particular interest, because the TUKH122 region
does not have any protostar
and shows the narrowest N$_2$H$^+$ line profiles found in our sample.
The non-thermal contribution to the CCS and N$_2$H$^+$ lines lies 
below the sonic value, implying that 
the cores are mainly thermally supported.
The observed CCS and N$_2$H$^+$ linewidth are 0.27$-$0.39 and 0.27$-$0.28 km s$^{-1}$,
respectively.
The thermal linewidth corresponding to 10 K for 
the mean molecular weight
(2.33 a.m.u.) is 0.44 km s$^{-1}$.
Then, the observed linewidth is subsonic.
When we adopt the definition of \citet{mye83},
the non-thermal (turbulent) linewidth $\Delta v$ (turb) is 0.25$-$0.38 and 0.24$-$0.25 km s$^{-1}$
for CCS and N$_2$H$^+$, respectively.

Figure 20 shows that there is no or a weak anticorrelation between $T_k$ and  $N$(N$_2$H$^+$)/$N$(CCS).
Some may think that $N$(N$_2$H$^+$)/$N$(CCS) has no correlation with $T_k$, while others
may think that $N$(N$_2$H$^+$)/$N$(CCS) is weakly anticorrelated with $T_k$.
Figure 19 shows similar relationship between $T_k$ and  $W$(N$_2$H$^+$)/$W$(CCS).
\citet{tat10} found that $N$(NH$_3$)/$N$(CCS) decreases with increasing
$T_{rot}$.
$N$(NH$_3$)/$N$(CCS) ranges from $\sim$3 to $\sim$100 in their study.
\citet{mar12} investigated the time evolution of the CCS and NH$_3$
abundances through the chemical model calculations with different kinetic
temperatures of 10, 15 and 25 K.
Their Figure 5 shows that the CCS abundance increases with increasing kinetic temperature,
at a given time.  The CCS abundance for 25 K is about a factor of 5 higher than that for 10 K,
for 1$-$3$\times$10$^4$ yr. However, For $>$ 3$\times$10$^4$ yr, the abundance difference
for 10$-$25 K becomes complicated.
The NH$_3$ abundance does not show simple increase or decrease with increasing kinetic temperature.
We wonder if there is a tendency that the CCS abundance increases with increasing kinetic temperature.
Figure 21 shows the NH$_3$ and CCS abundances against $T_K$ on the basis of \citet{wil99} and \citet{tat10}.
We used $N$(H$_2$) from C$^18$O 2$-$1 by \citet{wil99}.
The CCS abundance is almost constant against $T_k$, while the NH$_3$ abundance decreases with increasing
$T_k$.
When we use the CS column density from CS 1$-$0 as reference for abundance, we see similar tendencies
(not shown).
Therefore, observationally, we do not see evidence of $N$(CCS) increase with increasing $T_k$.
For warm regions ($T_{dust} \gtrsim$ 25 K), the abundance of N$_2$H$^+$ is expected to be lower for 
higher $T_k$ since CO evaporates from the mantles of dust grains.
The dust temperature $T_{dust}$ and the gas kinetic temperature $T_k$ can be different from each other, 
but may have some similarity.
Then, in warm region ($T_k \gtrsim$ 25 K), the N$_2$H$^+$ abundance can be decreased.
This could be a reason why Figure 20 shows that there can be a weak anticorrelation between 
$T_k$ and  $N$(N$_2$H$^+$)/$N$(CCS).

\citet{tat10} stated that the 
ratio of the column density
of NH$_3$ to that of CCS 
in the Orion A GMC is not necessarily high toward star-forming cores.
The column density ratio ranges from
6 to 200 when $T_{ex}$ (CCS) is assumed to be $T_{rot}$ (NH$_3$)/2.
Because the CCS $J_N$ = 4$_3-3_2$ observations by \citet{tat10} were single-point
observations toward the core center, and also
because the spatial resolution in CCS $J_N$ = 4$_3-3_2$ was twice worse than 
our current CCS $J_N$ = 7$_6-6_5$ observations,
the identification of starless and star-forming cores was
not accurate enough. 

\citet{sak06} and \citet{sak07} observed W3 GMC including AFGL 333 
in CCS $J_N$ = 4$_3-3_2$, NH$_3$, N$_2$H$^+$, 
and other lines.
They detected the CCS emission in two clumps in AFGL 333, and found that
$N$(N$_2$H$^+$)/$N$(CCS) is low (0.5$-$1.7)
in starless clump B and high (2$-$5)
star-forming clump A.
Our result based on a larger sample of cores is consistent with 
their result in W3 GMC.
\citet{sak08} observed IRDC (Infrared Dark Clouds) 
in CCS $J_N$ = 4$_3-3_2$, NH$_3$, N$_2$H$^+$, 
and other lines, and did not detect CCS $J_N$ = 4$_3-3_2$ in any objects.
The lower limit to $N$(N$_2$H$^+$)/$N$(CCS) in IRDCs is
mostly $\gtrsim$ 2,
judging from their sensitivity limit.
They concluded that IRDCs are more evolved than 
nearby dark clouds.
\citet{san12} studied IRDCs in N$_2$H$^+$, CCH and HC$_3$N, and other molecular species. 
They found that the total column densities of the different molecules, except CCH, increase with the evolutionary
stage of the clumps.
Their observations did not include CCS lines.  
According to the production mechanism of CCS proposed by \citet{suz92},
a reaction of CCH with S$^+$ is a
major route to produce CCS.
\citet{beu08} has shown that CCH is a tracer of young molecular gas.
Although CCS has not been observed in IRDCs yet, it can be just due to lower sensitivity.
\citet{tat10} compared their detection limit with that in \citet{sak08}.
Note that \citet{tat08} did not detect CCS $J_N$ = 4$_3-3_2$ but \citet{tat10} did
with better sensitivity, in the Orion A GMC.
The CCH column density does not show large variations among IRDCs in \citet{san12}.
If the production mechanism of CCS proposed by \citet{suz92} is valid for IRDCs, 
CCS is also expected to vary little across this IRDCs sample.
CCS observations toward IRDCs will let us compare the CCH and CCS abundance, and 
tell us how CCS forms in IRDCs.

According to \citet{ben98},
$N$(N$_2$H$^+$)/$N$(CCS)
ranges from 0.3 to 3.3 in dark clouds.
The column density ratio we derived ranges from
0.3 to 70, and the span is larger.
This may be due to the fact that Orion A GMC presents 
larger chemical variations than those seen in dark clouds, 
which can be
a consequence of the expected differences 
produced by larger variations in the kinetic temperature 
of the gas in the Orion A GMC (see \cite{mar12}).

\begin{figure}
  \begin{center}
    \FigureFile(150mm,150mm){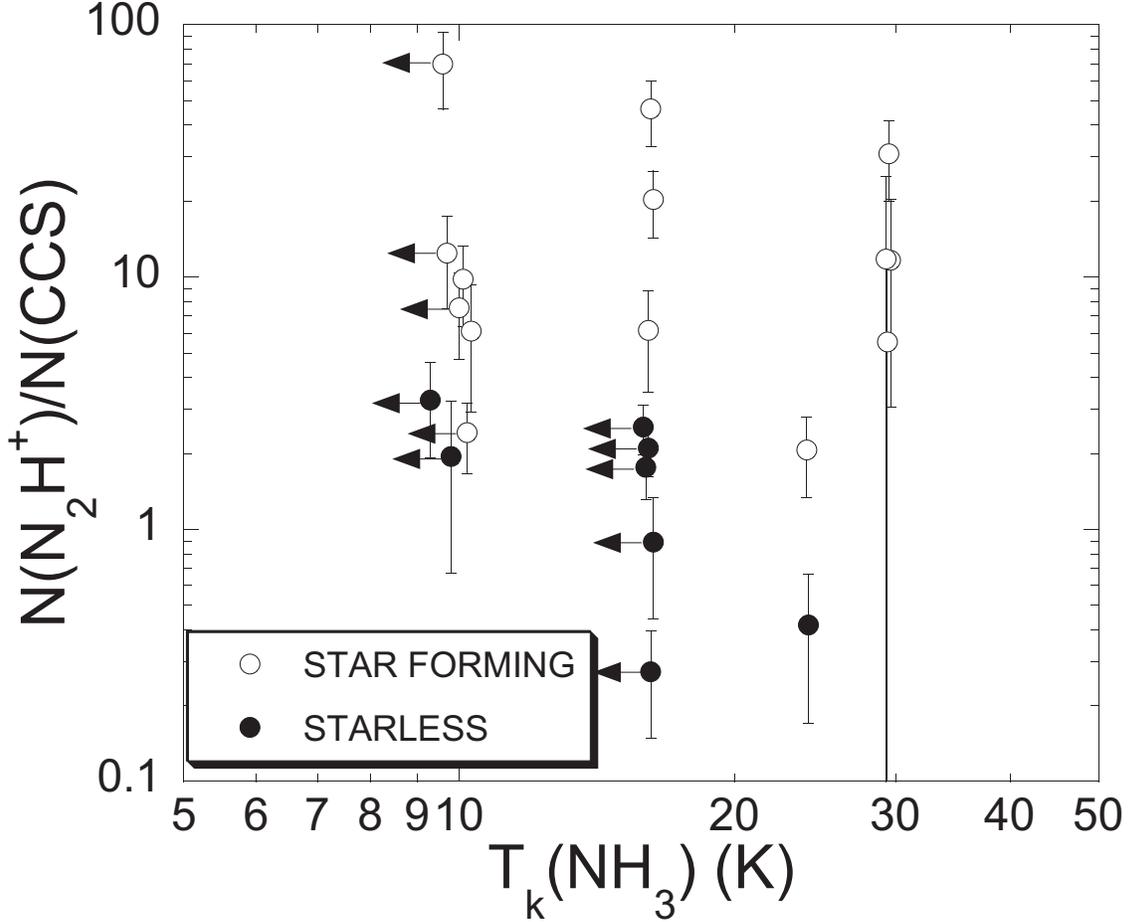}
  \end{center}
  \caption{The ratio of the column density
of N$_2$H$^+$ to that of CCS is
plotted against the gas kinetic temperature $T_k$. 
The error bar represent the 1$\sigma$ error in the optical depth only
in the N$_2$H$^+$
hyperfine line fitting.
}\label{fig:figure20}
\end{figure}

\begin{figure}
  \begin{center}
    \FigureFile(150mm,150mm){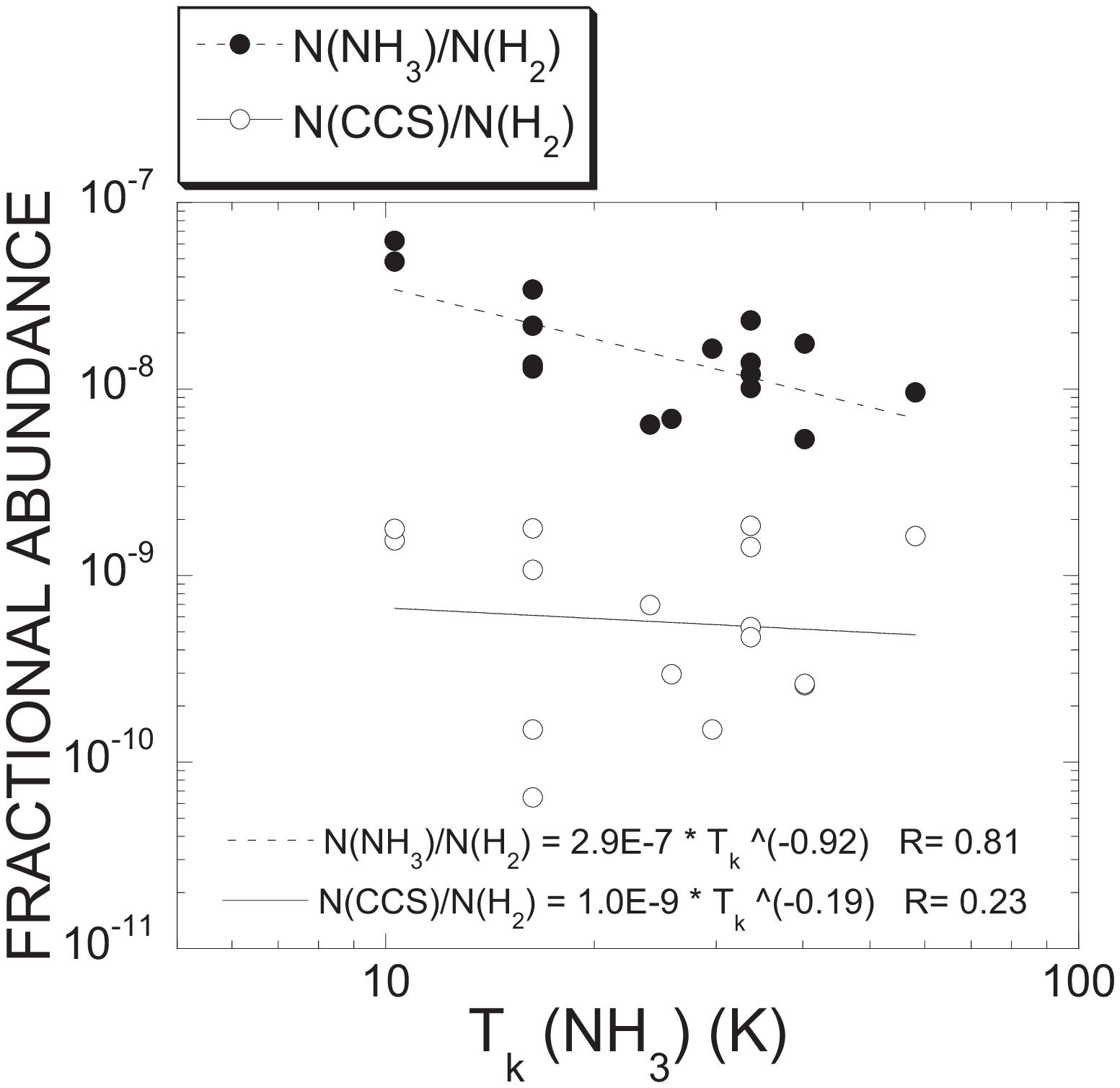}
  \end{center}
  \caption{Fractional abundances of NH$_3$ and CCS are plotted against the kinetic temperature $T_k$.
Fractional abundances are obtained from the ratio of column densities to N(H$_2$).
The NH$_3$ data and $N$(H$_2$), which was obtained from C$^{18}$O 2-1 data, are taken from \citet{wil99}.
The CCS column density is taken from \citet{tat10}, and here we adopt values obtained by assuming $T_{ex}$ = $T_{rot}$/2.
We used only cores which were detected in CCS $J_N$ = 4$_3-3_2$ and also have $T_{rot}$ measurements.
}\label{fig:figure21}
\end{figure}

\subsection{Validity of our results against the optical depth of the CCS lines}

To estimate a typical optical depth of the CCS $J_N$ = 7$_6-6_5$ emission,
we carry out the large velocity gradient (LVG) model again.
We take TUKH122 as an example having
low kinetic temperature, large molecular column density, and
narrow linewidth.
In CCS $J_N$ = 4$_3-3_2$ at 45.379033 GHz,
the intensity is $T_A^*$ = 0.47 K ($T_R$ = 0.64 K) and the linewidth
is 0.55 km s$^{-1}$.
In CCS $J_N$ = 7$_6-6_5$ at 81.505208 GHz,
the intensity is $T_A^*$ = 0.36 K ($T_R$ = 0.88 K) and the linewidth
is 0.27 km s$^{-1}$.
We assumed the common linewidth of 0.4 km s$^{-1}$ for LVG models.
Although the FWHM used in observations differs between
CCS $J_N$ = 4$_3-3_2$ at 45.379033 GHz (FWHM beam size = 39 arcsec)
and
CCS $J_N$ = 7$_6-6_5$ at 81.505208 GHz (19 arcsec),
we simply compare the intensities without any correction for different
beam sizes.
The observations and models are 
listed in Table 6.
``Observation'' represents the value obtained at TUKH122.
``Model'' represents the best fit models.
``(Model)'' represents models with the CCS column density obtained from the LTE calculation
for the CCS $J_N$ = 7$_6-6_5$ observations.
The derived column 
density of CCS by using the best-fit LVG model is a factor of 4$-$7 lower than that 
obtained from the CCS(7-6) line and by assuming LTE conditions.
When we take ``(Model)'', the absolute intensity at CCS $J_N$ = 4$_3-3_2$
tends to be higher than the value observed, although the intensity ratio is close to what was observed.
A possibility is that the beam filling factor is lower than unity.
Moreover, differences of the beam sizes at these two frequencies do not allow us to
make precise estimates of the intensity ratio.
Therefore, we will not go to details on the basis of LVG.
At least, we conclude that the optical depth of the CCS $J_N$ = 7$_6-6_5$ emission is thin or moderate.

\subsection{Variation of the N$_2$H$^+$/CCS ratio along the Orion A GMC}

\citet{tat10} found global variation along the Orion A GMC filament in the 
ratio of the column density
of NH$_3$ to that of CCS.
Figure 22 shows the 
ratio of the column density
of N$_2$H$^+$ to that of CCS is
plotted against declination.
Because the observed cores are distributed sparsely, it is hard to derive a tendency.
In the northern region (DEC $>$ $-$6.0 degrees), we only have star-forming peaks.

\begin{figure}
  \begin{center}
    \FigureFile(150mm,150mm){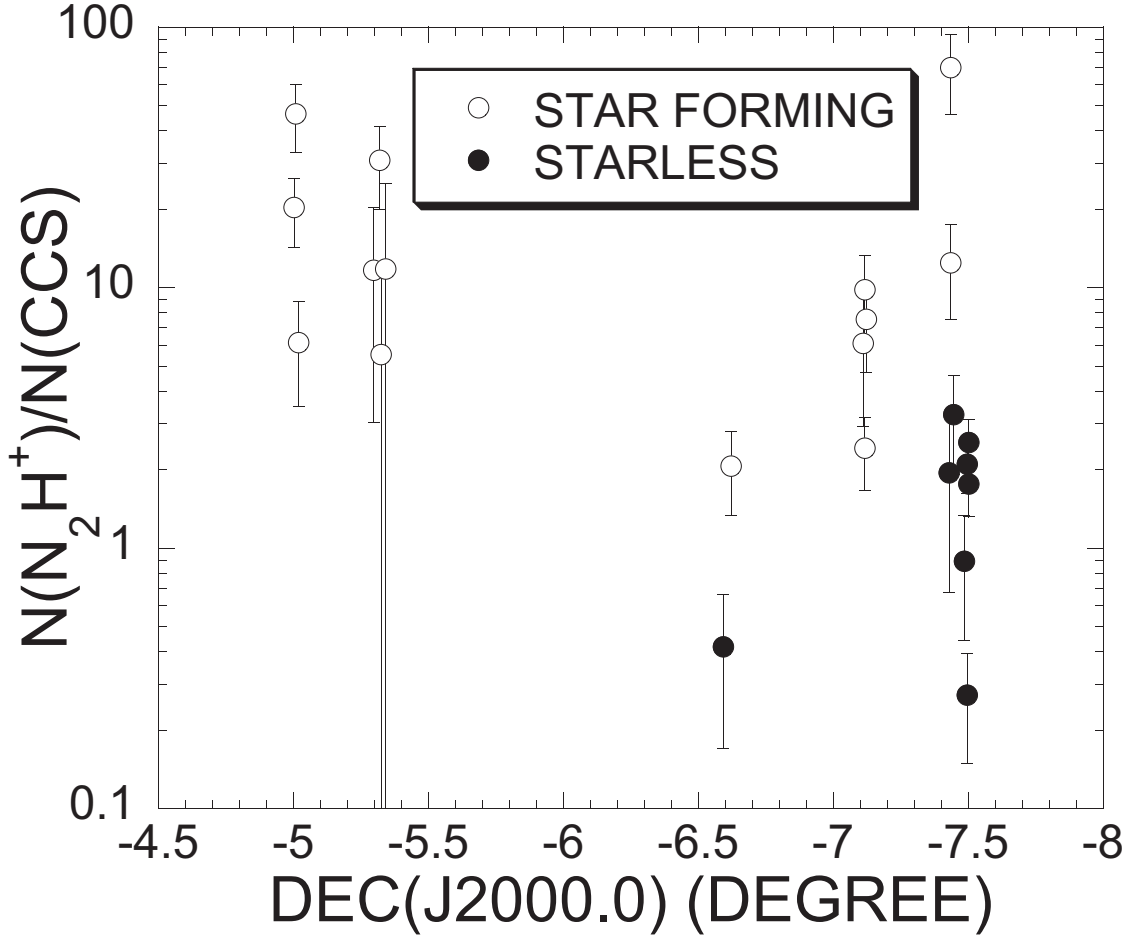}
  \end{center}
  \caption{The ratio of the column density
of N$_2$H$^+$ to that of CCS is
plotted against the declination. 
}\label{fig:figure22}
\end{figure}

\citet{tat10} did not detect the CCS $J_N$ = 4$_3-3_2$ emission at 45.379033GHz 
toward either Orion KL, which is very warm
($T_k$ = 70$-$150K: \cite{gen89}, \cite{wil99}), 
or the Orion Bar.
From Figure 5 of \citet{tat10}, the detection rate of the CCS $J_N$ = 4$_3-3_2$ emission at 45.379033GHz
is much lower (15$\%$) for $T_{rot} >$ 30 K ($T_k >$ 40 K) compared with 
that in the overall sample
(32$\%$).
\citet{tat08} did not detect the N$_2$H$^+$ $J$ = 1$-$0 emission 
toward the Orion Bar, the archetypal photodissociation region (PDR).
The N$_2$H$^+$ emission is very weak toward Orion KL.
It is known that N$_2$H$^+$ $J$ = 1$-$0 traces quiescent molecular gas, which is not affected by
star formation activity (e.g., \cite{wom93}).
It is possible that both the CCS and N$_2$H$^+$  $J$ = 1$-$0 
emission is weak in warm, evolved molecular gas.

It is expected that the evolution of clouds in nearby galaxies will become major
research topics in near future, because ALMA resolution and sensitivity will allow us 
to study it in great detail.
We believe that our finding on astrochemistry in the Orion A GMC 
will serve as a basic guide for astrochemical study of GMCs in
nearby galaxies by ALMA.

\section{Summary} 

We mapped cores in the Orion A GMC
($T_k \sim$ 10$-$30 K)
in CCS $J_N$ = 7$_6-6_5$ and N$_2$H$^+$.
It is found for cores with temperatures lower than 25 K 
that 
the CCS $J_N$ = 7$_6-6_5$ emission tends to be strong in starless peaks
while
the N$_2$H$^+$ $J$ = 1$-$0 $F_1$ = 2$-$1 emission tends to
be strong toward protostars.
In some cases, we detected local weak CCS $J_N$ = 7$_6-6_5$ peaks toward protostars as well as 
N$_2$H$^+$ $J$ = 1$-$0 $F_1$ = 2$-$1 peaks.
This may be caused by secondary late-stage CCS peak in the chemical evolution.
The column density ratio of $N$(N$_2$H$^+$)/$N$(CCS) is
low toward starless peaks while it is high
toward star-forming peaks.
This tendency is consistent with the tendency found 
in dark clouds ($T_k \sim$ 10 K).
The criterion found in the Orion A GMC is $N$(N$_2$H$^+$)/$N$(CCS) $\sim 2-3$.
On the other hand, some protostars do not accompany any
N$_2$H$^+$ $J$ = 1$-$0 $F_1$ = 2$-$1 intensity peaks but are associated with dust continuum emitting regions, 
suggesting that the N$_2$H$^+$ abundance 
might be decreased 
due to CO evaporation in warmer star-forming sites.

\bigskip

The authors would like to thank an anonymous referee, whose comments greatly improved the paper. 
J.-E.L. was supported by the Basic Science Research Program through the National Research Foundation of Korea (NRF) funded by the Ministry of Education of the Korean government (grant number NRF-2012R1A1A2044689) and the 2013 Sabbatical Leave Program of Kyung Hee University (KHU-20131724).  
M.C. was supported by
the Core Research Program of National Research Foundation
funded by the Ministry of Science, ICT and Future Planning
of the Korean government (grant number NRF-2011-0015816).


\begin{longtable}{llllll}
  \caption{Map Centers and Off Positions}\label{tab:LTsample}
  \hline              

TUKH	&	Map Center	&		&	Off Position	&		&	rms (CCS) at 30.52 kHz	\\
	&	RA(J2000.0)	&	DEC(J2000.0)	&	RA(J2000.0)	&	DEC(J2000.0)	&		\\
  \hline  
	&	h m s	&	$\degree$  $\arcmin$  $\arcsec$	&	h m s	&	$\degree$  $\arcmin$  $\arcsec$	&	K	\\

\endfirsthead
\hline
\endhead
  \hline
\endfoot
  \hline
\endlastfoot
  \hline

003	&	5:35:17.7	&	-5:00:30	&	5:33:17.2	&	-5:00:30	&	0.037 	\\
021	&	5:35:17.2	&	-5:19:10	&	5:33:16.7	&	-5:19:10	&	0.060 	\\
040	&	5:35:11.7	&	-5:29:09	&	5:33:11.1	&	-5:29:09	&	0.075 	\\
049	&	5:34:50.1	&	-5:39:08	&	5:32:49.5	&	-5:39:08	&	0.072 	\\
056	&	5:35:11.1	&	-5:55:49	&	5:33:10.5	&	-5:55:49	&	0.051 	\\
059	&	5:35:40.4	&	-6:07:52	&	5:33:39.7	&	-6:07:52	&	0.051 	\\
069	&	5:36:25.8	&	-6:23:15	&	5:34:25.1	&	-6:23:15	&	0.076 	\\
083	&	5:36:47.0	&	-6:31:56	&	5:34:46.2	&	-6:31:56	&	0.070 	\\
088	&	5:37:00.4	&	-6:35:57	&	5:34:59.6	&	-6:35:57	&	0.057 	\\
097	&	5:37:59.0	&	-7:07:22	&	5:35:58.1	&	-7:07:22	&	0.048 	\\
104	&	5:39:06.2	&	-7:12:07	&	5:37:05.2	&	-7:12:07	&	0.051 	\\
105	&	5:38:07.0	&	-7:13:22	&	5:36:06.0	&	-7:13:22	&	0.050 	\\
117	&	5:39:18.4	&	-7:26:47	&	5:37:17.4	&	-7:26:47	&	0.052 	\\
122	&	5:39:42.5	&	-7:30:09	&	5:37:41.5	&	-7:30:09	&	0.052 	\\

\end{longtable}

\begin{longtable}{lllllllll}
  \caption{Properties of the Young Stellar Objects}\label{tab:LTsample}
  \hline              
Megeath	&	RA(J2000.0)	&	DEC(J2000.0)	&	$T_{bol}$	&	Class	&	$\alpha_{IRAC}$	&	$L_{bol}$	&	Other Name	&	Reference for Other Name	\\
  \hline  
	&	h m s	&	$\degree$  $\arcmin$  $\arcsec$	&	K	&		&		&	$L_{\odot}$	&		&		\\
\endfirsthead
\hline
\endhead
  \hline
\endfoot
  \hline
\endlastfoot
  \hline

545	&	5:39:19.61	&	-7:26:18.8	&	505.3 	&	P	&	0.34	&	1.32 	&	Haro 4-255 VLA1	&	\citet{ang92}	\\
551	&	5:39:19.98	&	-7:26:11.2	&	179.2 	&	P	&	1.18	&	1.23 	&	Haro 4-255 FIR	&	\citet{eva86}	\\
638	&	5:37:58.76	&	-7: 7:25.3	&	176.6 	&	P	&	2.49	&	0.08 	&		&		\\
640	&	5:37:57.01	&	-7: 6:56.5	&	182.0 	&	P	&	1.45	&	0.11 	&		&		\\
826	&	5:37:00.45	&	-6:37:10.5	&	643.8 	&	FP	&	-1	&	0.02 	&		&		\\
1968	&	5:35:14.93	&	-5:21: 0.6	&	1344.3 	&	P	&		&	0.08 	&		&		\\
1969	&	5:35:11.84	&	-5:21: 0.3	&	716.3 	&	P	&	-0.15	&	0.94 	&		&		\\
1971	&	5:35:17.84	&	-5:20:53.9	&	1778.1 	&	P	&		&	1.11 	&		&		\\
2017	&	5:35:13.60	&	-5:19:54.9	&	1245.6 	&	P	&	0.06	&	4.44 	&		&		\\
2037	&	5:35:15.39	&	-5:19:34.4	&	771.4 	&	P	&		&	0.02 	&		&		\\
2044	&	5:35:20.71	&	-5:19:26.3	&	780.2 	&	P	&	0.95	&	0.81 	&		&		\\
2050	&	5:35:11.61	&	-5:19:12.4	&	1407.9 	&	P	&		&	0.09 	&		&		\\
2060	&	5:35:22.10	&	-5:18:57.7	&	921.5 	&	P	&		&	0.06 	&		&		\\
2069	&	5:35:16.69	&	-5:18:45.2	&	788.5 	&	P	&	0.29	&	0.25 	&		&		\\
2073	&	5:35:12.02	&	-5:18:40.8	&	961.3 	&	P	&		&	0.03 	&		&		\\
2091	&	5:35:17.09	&	-5:18:13.9	&	893.8 	&	P	&		&	0.04 	&		&		\\
2106	&	5:35:19.66	&	-5:17:46.2	&	779.1 	&	P	&		&	0.03 	&		&		\\
2427	&	5:35:23.65	&	-5: 1:40.3	&	246.2 	&	P	&	2.8	&	0.90 	&	SMM9	&	\citet{tak13}	\\
2433	&	5:35:23.47	&	-5: 1:28.7	&	176.9 	&	P	&	2.35	&	0.57 	&	MMS6-NE	&	\citet{tak09}	\\
	&		&		&		&		&		&		&	SMM7	&	\citet{tak13}	\\
2437	&	5:35:22.43	&	-5: 1:14.1	&	161.6 	&	P	&	1.76	&	0.36 	&	MMS5	&	\citet{chi97}	\\
	&		&		&		&		&		&		&	SMM6	&	\citet{tak13}	\\
2440	&	5:35:19.96	&	-5: 1: 2.6	&	483.7 	&	P	&	0.4	&	0.42 	&		&		\\
2442	&	5:35:18.91	&	-5: 0:50.9	&	220.8 	&	P	&		&	0.18 	&	TKH10	&	\citet{tsu01}	\\
	&		&		&		&		&		&		&	MMS3	&	\citet{chi97}	\\
	&		&		&		&		&		&		&	SMM4	&	\citet{tak13}	\\
2446	&	5:35:18.32	&	-5: 0:33.0	&	455.7 	&	P	&	0.77	&	6.55 	&	TKH8	&	\citet{tsu01}	\\
	&		&		&		&		&		&		&	MMS2	&	\citet{chi97}	\\
	&		&		&		&		&		&		&	SMM3	&	\citet{tak13}	\\
2451	&	5:35:15.03	&	-5: 0: 8.2	&	272.8 	&	P	&	-0.16	&	0.10 	&		&		\\
2453	&	5:35:16.15	&	-5: 0: 2.3	&	324.6 	&	P	&	0.39	&	1.86 	&	CSO3	&	\citet{lis98}	\\
	&		&		&		&		&		&		&	SMM1	&	\citet{tak13}	\\

\end{longtable}



\begin{longtable}{llllllll}
  \caption{Intensity Peak, Coordinate, Kinetic Temperature from NH$_3$, and CCS Line Parameter}\label{tab:LTsample}
  \hline              

TUKH	&	RA(J2000.0)	&	DEC(J2000.0)	&	$T_k$	&		$T_A^*$	&	LSR Velocity	&	$\Delta v$	&	rms at 122.08 kHz	\\
  \hline    
	&	h m s	&	$\degree$  $\arcmin$  $\arcsec$	&		K	&	K	&	km s$^{-1}$	&	km s$^{-1}$	&	K	\\

\endfirsthead
\hline
\endhead
  \hline
\endfoot
  \hline
\endlastfoot
  \hline

003A	&	5:35:17.7	&	-5:00:10	&	16 	&	0.09 	$\pm$	0.02	&	10.53 	$\pm$	0.13	&	1.69 	$\pm$	0.32 	&	0.02 	\\
003B	&	5:35:19.0	&	-5:00:30	&	16 	&	0.09 	$\pm$	0.02	&	10.98 	$\pm$	0.11	&	1.18 	$\pm$	0.28 	&	0.02 	\\
003C	&	5:35:17.7	&	-5:01:10	&	16 	&	0.14 	$\pm$	0.02	&	10.54 	$\pm$	0.05	&	0.75 	$\pm$	0.13 	&	0.02 	\\
021A	&	5:35:19.9	&	-5:17:50	&	30 	&	0.10 	$\pm$	0.02	&	9.23 	$\pm$	0.23	&	2.65 	$\pm$	0.55 	&	0.03 	\\
021B	&	5:35:21.2	&	-5:18:50	&	30 	&	0.15 	$\pm$	0.03	&	8.53 	$\pm$	0.06	&	0.54 	$\pm$	0.15 	&	0.03 	\\
021	&	5:35:17.2	&	-5:19:10	&	30 	&	0.30 	$\pm$	0.02	&	9.30 	$\pm$	0.05	&	1.39 	$\pm$	0.12 	&	0.02 	\\
021C	&	5:35:14.5	&	-5:19:30	&	30 	&	0.21 	$\pm$	0.03	&	9.03 	$\pm$	0.11	&	1.78 	$\pm$	0.25 	&	0.04 	\\
021D	&	5:35:13.2	&	-5:19:50	&	30 	&	0.16 	$\pm$	0.02	&	9.15 	$\pm$	0.15	&	2.17 	$\pm$	0.35 	&	0.04 	\\
021E	&	5:35:15.9	&	-5:20:30	&	30 	&	0.17 	$\pm$	0.01	&	9.77 	$\pm$	0.15	&	2.71 	$\pm$	0.36 	&	0.04 	\\
088A	&	5:37:00.4	&	-6:35:37	&	24 	&	0.10 	$\pm$	0.02	&	6.50 	$\pm$	0.13	&	1.42 	$\pm$	0.30 	&	0.03 	\\
088B	&	5:37:00.4	&	-6:37:17	&	24 	&	0.27 	$\pm$	0.03	&	5.97 	$\pm$	0.05	&	0.78 	$\pm$	0.11 	&	0.03 	\\
088C	&	5:36:59.1	&	-6:37:57	&	24 	&	0.13 	$\pm$	0.03	&	5.79 	$\pm$	0.12	&	1.06 	$\pm$	0.30 	&	0.04 	\\
097A	&	5:37:57.7	&	-7:06:42	&	10 	&	0.16 	$\pm$	0.02	&	5.80 	$\pm$	0.07	&	1.39 	$\pm$	0.17 	&	0.04 	\\
097B	&	5:37:56.3	&	-7:07:02	&	10 	&	0.19 	$\pm$	0.03	&	6.11 	$\pm$	0.07	&	0.87 	$\pm$	0.16 	&	0.04 	\\
097C	&	5:37:57.7	&	-7:07:02	&	10 	&	0.19 	$\pm$	0.02	&	6.14 	$\pm$	0.04	&	0.75 	$\pm$	0.10 	&	0.02 	\\
097D	&	5:38:01.7	&	-7:07:02	&	10 	&	0.12 	$\pm$	0.02	&	5.41 	$\pm$	0.08	&	0.96 	$\pm$	0.18 	&	0.04 	\\
097	&	5:37:59.0	&	-7:07:22	&	10 	&	0.16 	$\pm$	0.02	&	5.95 	$\pm$	0.06	&	0.87 	$\pm$	0.14 	&	0.02 	\\
097E	&	5:38:00.3	&	-7:08:02	&	10 	&	0.15 	$\pm$	0.04	&	6.14 	$\pm$	0.05	&	0.38 	$\pm$	0.12 	&	0.03 	\\
117A	&	5:39:22.4	&	-7:25:47	&	$<$10	&	0.20 	$\pm$	0.02	&	4.22 	$\pm$	0.04	&	0.72 	$\pm$	0.09 	&	0.03 	\\
117B	&	5:39:18.4	&	-7:26:07	&	$<$10	&	0.15 	$\pm$	0.03	&	4.07 	$\pm$	0.05	&	0.51 	$\pm$	0.12 	&	0.03 	\\
117C	&	5:39:19.7	&	-7:26:07	&	$<$10	&	0.09 	$\pm$	0.04	&	4.43 	$\pm$	0.09	&	0.57 	$\pm$	0.21 	&	0.02 	\\
117D	&	5:39:17.1	&	-7:26:27	&	$<$10	&	0.10 	$\pm$	0.02	&	3.88 	$\pm$	0.08	&	0.81 	$\pm$	0.18 	&	0.02 	\\
117E	&	5:39:22.4	&	-7:26:27	&	$<$10	&	0.20 	$\pm$	0.03	&	4.19 	$\pm$	0.07	&	0.87 	$\pm$	0.19 	&	0.04 	\\
117F	&	5:39:21.1	&	-7:26:47	&	$<$10	&	0.19 	$\pm$	0.03	&	4.34 	$\pm$	0.04	&	0.69 	$\pm$	0.10 	&	0.03 	\\
117G	&	5:39:19.7	&	-7:27:07	&	$<$10	&	0.08 	$\pm$	0.01	&	4.48 	$\pm$	0.11	&	1.36 	$\pm$	0.29 	&	0.02 	\\
122A	&	5:39:41.2	&	-7:29:09	&	$<$16	&	0.19 	$\pm$	0.03	&	3.82 	$\pm$	0.05	&	0.57 	$\pm$	0.12 	&	0.03 	\\
122B	&	5:39:41.2	&	-7:29:49	&	$<$16	&	0.22 	$\pm$	0.03	&	3.75 	$\pm$	0.02	&	0.36 	$\pm$	0.05 	&	0.01 	\\
122C	&	5:39:42.5	&	-7:29:49	&	$<$16	&	0.24 	$\pm$	0.04	&	3.78 	$\pm$	0.03	&	0.39 	$\pm$	0.08 	&	0.02 	\\
122	&	5:39:42.5	&	-7:30:09	&	$<$16	&	0.36 	$\pm$	0.03	&	3.74 	$\pm$	0.01	&	0.27 	$\pm$	0.03 	&	0.02 	\\
122D	&	5:39:43.8	&	-7:30:09	&	$<$16	&	0.19 	$\pm$	0.03	&	3.75 	$\pm$	0.03	&	0.39 	$\pm$	0.07 	&	0.02 	\\
122E	&	5:39:46.5	&	-7:30:49	&	$<$16	&	0.26 	$\pm$	0.07	&	3.82 	$\pm$	0.03	&	0.21 	$\pm$	0.06 	&	0.03 	\\

\end{longtable}

\begin{longtable}{llllll}
  \caption{N$_2$H$^+$ Velocity-Integrated Intensity $W$ and Hyperfine Line Fitting}\label{tab:LTsample}
  \hline              
TUKH	&	$W$	&	$\tau_{TOT}$			&	LSR Velocity			&	$\Delta v$			&	T$_{ex}$			\\
  \hline 
	&	K km s$^{-1}$	&				&	km s$^{-1}$			&	km s$^{-1}$			&	K			\\

\endfirsthead
\hline
\endhead
  \hline
\endfoot
  \hline
\endlastfoot
  \hline
003A	&	5.6 	&	5.9 	$\pm$	1.1 	&	11.18 	$\pm$	0.01 	&	0.77 	$\pm$	0.04 	&	9.4 	$\pm$	0.6 	\\
003B	&	5.8 	&	5.7 	$\pm$	1.1 	&	11.05 	$\pm$	0.01 	&	0.81 	$\pm$	0.04 	&	11.2 	$\pm$	0.7 	\\
003C	&	1.8 	&	3.7 	$\pm$	1.1 	&	10.46 	$\pm$	0.01 	&	0.51 	$\pm$	0.02 	&	7.1 	$\pm$	0.7 	\\
021A	&	5.8 	&	1.4 	$\pm$	0.6 	&	9.29 	$\pm$	0.02 	&	1.46 	$\pm$	0.07 	&	14.1 	$\pm$	3.8 	\\
021B	&	0.8 	&				&	9.36 	$\pm$	0.05 	&				&				\\
021	&	12.5 	&	3.3 	$\pm$	0.7 	&	9.73 	$\pm$	0.02 	&	1.30 	$\pm$	0.06 	&	20.8 	$\pm$	1.8 	\\
021C	&	4.0 	&	0.7 	$\pm$	0.5 	&	9.44 	$\pm$	0.03 	&	2.00 	$\pm$	0.09 	&	13.6 	$\pm$	6.9 	\\
021D	&	1.8 	&				&	9.41 	$\pm$	0.03 	&	1.54 	$\pm$	0.11 	&				\\
021E	&	8.2 	&	1.0 	$\pm$	0.6 	&	9.94 	$\pm$	0.02 	&	1.29 	$\pm$	0.06 	&	27.3 	$\pm$	12.7 	\\
088A	&	0.5 	&	5.4 	$\pm$	2.4 	&	6.53 	$\pm$	0.03 	&	0.81 	$\pm$	0.09 	&	3.4 	$\pm$	0.1 	\\
088B	&	1.4 	&	5.4 	$\pm$	1.3 	&	5.80 	$\pm$	0.02 	&	0.88 	$\pm$	0.06 	&	4.8 	$\pm$	0.2 	\\
088C	&	0.1 	&				&				&				&				\\
097A	&	2.5 	&	2.9 	$\pm$	1.0 	&	5.69 	$\pm$	0.02 	&	0.82 	$\pm$	0.04 	&	8.5 	$\pm$	1.1 	\\
097B	&	1.5 	&	6.9 	$\pm$	1.4 	&	6.04 	$\pm$	0.01 	&	0.41 	$\pm$	0.02 	&	5.4 	$\pm$	0.3 	\\
097C	&	2.8 	&	4.2 	$\pm$	1.0 	&	5.86 	$\pm$	0.02 	&	0.84 	$\pm$	0.04 	&	7.5 	$\pm$	0.5 	\\
097D	&	0.7 	&				&				&				&				\\
097	&	2.1 	&	3.9 	$\pm$	1.0 	&	5.77 	$\pm$	0.01 	&	0.77 	$\pm$	0.04 	&	7.1 	$\pm$	0.6 	\\
097E	&	0.7 	&	2.5 	$\pm$	1.8 	&	6.12 	$\pm$	0.02 	&	0.56 	$\pm$	0.04 	&	4.6 	$\pm$	1.0 	\\
117A	&	1.4 	&	3.5 	$\pm$	1.5 	&	4.50 	$\pm$	0.02 	&	0.65 	$\pm$	0.04 	&	5.2 	$\pm$	0.8 	\\
117B	&	2.9 	&	3.9 	$\pm$	1.0 	&	4.76 	$\pm$	0.02 	&	0.81 	$\pm$	0.04 	&	6.7 	$\pm$	0.6 	\\
117C	&	5.1 	&	4.5 	$\pm$	1.0 	&	4.65 	$\pm$	0.01 	&	0.81 	$\pm$	0.04 	&	11.0 	$\pm$	0.8 	\\
117D	&	0.6 	&				&	4.55 	$\pm$	0.04 	&	1.00 	$\pm$	0.12 	&	4.6 	$\pm$	3.0 	\\
117E	&	1.6 	&				&				&				&				\\
117F	&	1.4 	&	4.1 	$\pm$	1.2 	&	4.42 	$\pm$	0.01 	&	0.66 	$\pm$	0.03 	&	5.6 	$\pm$	0.4 	\\
117G	&	0.4 	&	3.1 	$\pm$	2.6 	&	4.38 	$\pm$	0.02 	&	0.53 	$\pm$	0.05 	&	3.8 	$\pm$	0.6 	\\
122A	&	0.6 	&	6.2 	$\pm$	2.3 	&	3.95 	$\pm$	0.01 	&	0.41 	$\pm$	0.03 	&	4.0 	$\pm$	0.2 	\\
122B	&	0.4 	&	7.8 	$\pm$	2.7 	&	3.85 	$\pm$	0.01 	&	0.42 	$\pm$	0.03 	&	3.4 	$\pm$	0.1 	\\
122C	&	0.9 	&	28.2 	$\pm$	5.1 	&	3.82 	$\pm$	0.01 	&	0.28 	$\pm$	0.01 	&	3.8 	$\pm$	0.0 	\\
122	&	0.8 	&	21.8 	$\pm$	4.2 	&	3.80 	$\pm$	0.01 	&	0.28 	$\pm$	0.01 	&	4.1 	$\pm$	0.1 	\\
122D	&	0.8 	&	20.3 	$\pm$	3.4 	&	3.78 	$\pm$	0.01 	&	0.27 	$\pm$	0.01 	&	4.0 	$\pm$	0.1 	\\
122E	&	0.5 	&				&				&				&				\\

\end{longtable}


\begin{longtable}{lllcll}
  \caption{Column Density and Ratios}\label{tab:LTsample}
  \hline              

TUKH	&	$N$(CCS)	&	$N$(N$_2$H$^+$)			&	$W$(N$_2$H$^+$)/$W$(CCS)	&	$N$(N$_2$H$^+$)/$N$(CCS)			&	Megeath	\\
\hline
	&	cm$^{-2}$	&	cm$^{-2}$			&		&				&		\\

\hline
\endhead
  \hline
\endfoot
  \hline
\endlastfoot
  \hline
003A	&	2.7E+12	&	5.6E+13	$\pm$	1.0E+13	&	33.0 	&	20.3 	$\pm$	6.0 	&	2453, 2446	\\
003B	&	1.5E+12	&	7.1E+13	$\pm$	1.3E+13	&	54.1 	&	46.5 	$\pm$	13.5 	&	2440, 2442, 2446	\\
003C	&	2.6E+12	&	1.6E+13	$\pm$	4.7E+12	&	15.8 	&	6.2 	$\pm$	2.7 	&	2442	\\
021A	&	3.7E+12	&	4.3E+13	$\pm$	1.8E+13	&	20.4 	&	11.7 	$\pm$	8.6 	&	2106	\\
021B	&		&				&	9.5 	&				&	2060	\\
021	&	5.7E+12	&	1.8E+14	$\pm$	3.8E+13	&	28.8 	&	30.8 	$\pm$	10.8 	&	2069	\\
021C	&	5.2E+12	&	2.9E+13	$\pm$	2.1E+13	&	10.2 	&	5.5 	$\pm$	7.1 	&	2037	\\
021D	&		&				&	4.9 	&				&	2017	\\
021E	&	6.9E+12	&	8.1E+13	$\pm$	4.9E+13	&	16.7 	&	11.8 	$\pm$	13.2 	&	1968	\\
088A	&	5.3E+13	&	2.2E+13	$\pm$	1.0E+13	&	3.5 	&	0.4 	$\pm$	0.2 	&		\\
088B	&	1.4E+13	&	2.9E+13	$\pm$	7.2E+12	&	6.5 	&	2.1 	$\pm$	0.7 	&	826	\\
088C	&	2.0E+12	&				&	0.6 	&				&		\\
097A	&	4.2E+12	&	2.6E+13	$\pm$	8.7E+12	&	10.8 	&	6.1 	$\pm$	3.2 	&	640	\\
097B	&	7.7E+12	&	1.9E+13	$\pm$	3.9E+12	&	8.6 	&	2.4 	$\pm$	0.8 	&	640	\\
097C	&	3.3E+12	&	3.3E+13	$\pm$	7.7E+12	&	17.7 	&	9.8 	$\pm$	3.5 	&	640	\\
097D	&	5.7E+12	&				&	5.9 	&				&		\\
097	&	3.5E+12	&	2.6E+13	$\pm$	6.6E+12	&	14.5 	&	7.6 	$\pm$	2.8 	&	638	\\
097E	&	4.3E+12	&	8.1E+12	$\pm$	5.8E+12	&	11.5 	&	1.9 	$\pm$	1.9 	&		\\
117A	&	7.4E+12	&	1.4E+13	$\pm$	6.4E+12	&	8.9 	&	1.9 	$\pm$	1.3 	&		\\
117B	&	2.0E+12	&	2.6E+13	$\pm$	6.5E+12	&	36.0 	&	12.5 	$\pm$	4.9 	&	545, 551	\\
117C	&	8.0E+11	&	5.6E+13	$\pm$	1.2E+13	&	91.1 	&	70.0 	$\pm$	23.6 	&	545, 551	\\
117D	&	6.0E+12	&				&	6.5 	&				&		\\
117E	&	9.0E+12	&				&	8.6 	&				&		\\
117F	&	5.6E+12	&	1.8E+13	$\pm$	5.3E+12	&	9.4 	&	3.3 	$\pm$	1.3 	&		\\
117G	&	1.8E+13	&	8.8E+12	$\pm$	7.4E+12	&	3.0 	&	0.5 	$\pm$	0.5 	&		\\
122A	&	1.5E+13	&	1.4E+13	$\pm$	5.1E+12	&	5.7 	&	0.9 	$\pm$	0.4 	&		\\
122B	&	6.2E+13	&	1.7E+13	$\pm$	5.8E+12	&	4.5 	&	0.3 	$\pm$	0.1 	&		\\
122C	&	2.0E+13	&	4.2E+13	$\pm$	7.6E+12	&	8.9 	&	2.1 	$\pm$	0.5 	&		\\
122	&	1.9E+13	&	3.4E+13	$\pm$	6.5E+12	&	7.6 	&	1.8 	$\pm$	0.4 	&		\\
122D	&	1.2E+13	&	3.0E+13	$\pm$	5.0E+12	&	9.9 	&	2.5 	$\pm$	0.6 	&		\\
122E	&	1.2E+12	&				&		&				&		\\

\end{longtable}

\begin{longtable}{cccccccc}
  \caption{Best-fit LVG Models for CCS toward TUKH122}\label{tab:LTsample}
  \hline 

Observation or Model & $n$  & $T_k$ & $N$ (CCS) & $T_R$ (4$_3-3_2$)& $T_R$ (7$_6-6_5$)$/$$T_R$ (4$_3-3_2$) & $\tau$ (4$_3-3_2$) & $\tau$ (7$_6-6_5$) \\  
  \hline          
 &cm$^{-3}$ & K & cm$^{-2}$ & K & & & \\    
\hline
\endhead
  \hline
\endfoot
  \hline
\endlastfoot
  \hline
Observation & ...  & ... & ...   & 0.64 & 1.38 & ...   & ... \\
Model       & 1E6 & 10 & 5E12 & 0.78 & 1.18 & 0.12 & 0.15\\
Model       & 3E6 & 10 & 5E12 & 0.77 & 1.21 & 0.11 & 0.15\\
Model       & 1E5 & 15 & 5E12 & 0.83 & 1.24 & 0.06 & 0.16\\
Model       & 1E5 & 20 & 3E12 & 0.45 & 1.55 & 0.02 & 0.08\\

(Model)       & 1E5 & 10 & 2E13 & 3.0  & 0.83 & 0.53 & 0.77\\
(Model)       & 1E5 & 20 & 2E13 & 3.8  & 1.41 & 0.14 & 0.49\\

\end{longtable}


\end{document}